\documentclass[english,prd,twocolumn,superscriptaddress,floatfix,nofootinbib,preprintnumbers,eqsecnum]{revtex4-1}
\usepackage{graphicx}
\usepackage{amsmath,amsthm,amssymb}
\usepackage{bm}

\usepackage{amsfonts}
\usepackage{dcolumn}
\usepackage{hyperref}

\def\be{\begin{equation}}
\def\ee{\end{equation}}
\def\ba{\begin{eqnarray}}
\def\ea{\end{eqnarray}}
\def\bs{\begin{subequations}}
\def\es{\end{subequations}}

\usepackage{color}

\makeatother

\usepackage{babel}
\makeatother

\begin{document}

\title{Cosmology with a successful Vainshtein screening 
in theories beyond Horndeski}

\author{Ryotaro Kase}

\affiliation{Department of Physics, Faculty of Science,
Tokyo University of Science, 1-3,
Kagurazaka, Shinjuku, Tokyo 162-8601, Japan}

\author{Shinji Tsujikawa}

\affiliation{Department of Physics, Faculty of Science, 
Tokyo University of Science, 1-3,
Kagurazaka, Shinjuku, Tokyo 162-8601, Japan}

\author{Antonio De Felice}

\affiliation{Yukawa Institute for Theoretical Physics,
Kyoto University, 606-8502, Kyoto, Japan}

\begin{abstract}

We propose a dark energy model where a scalar field $\phi$ 
has nonlinear self-interactions in the presence 
of a dilatonic coupling with the Ricci scalar. 
This belongs to a sub-class of theories beyond Horndeski, 
which accommodates covariant Galileons and Brans-Dicke 
theories as specific cases.
We derive conditions under which the scalar sound speed 
squared $c_{\rm s}^2$ is positive from the radiation era
to today. Since $c_{\rm s}^2$ remains to be smaller than 
the order of 1, the deviation from Horndeski theories does not 
cause heavy oscillations of gauge-invariant gravitational 
potentials. In this case, the evolution of matter perturbations 
at low redshifts is similar to that in the coupled dark energy 
scenario with an enhanced gravitational interaction. 
On the spherically symmetric background with a matter source, 
the existence of field self-interactions suppresses the 
propagation of fifth force inside a Vainshtein radius. 
We estimate an allowed parameter space in which the model 
can be compatible with Solar System constraints while satisfying 
conditions for the cosmological viability of background and perturbations.

\end{abstract}

\preprint{YITP-15-88}

\date{\today}

%\pacs{98.80.Cq}

\maketitle

%============================================
%section I
\section{Introduction}
\label{intro} 
%===========================================

The constantly accumulating observational evidence for 
a late-time acceleration of the Universe \cite{SNIa,CMB,BAO} 
implies that the origin of dark energy may be attributed to 
some scalar field degree of freedom or some modification 
of gravity from General Relativity (GR) \cite{CST}.
A canonical scalar field dubbed quintessence 
can drive the cosmic acceleration in the presence 
of a slowly varying potential \cite{quin1,quin2}.
In modified gravitational theories, there exists also 
a dynamical scalar degree of freedom arising from 
the breaking of gauge symmetries 
of GR \cite{moreview,fRreview}.

If we try to construct dark energy models in the framework of 
supersymmetric theories like string theory, there are in general 
couplings between a scalar field and the gravity sector \cite{darkco}.
The string dilaton $\phi$, for example, is coupled to the 
Ricci scalar $R$ in the form $F(\phi)R$ with 
$F(\phi)=e^{-2q\phi/M_{\rm pl}}$, 
where $q$ is a constant and $M_{\rm pl}$ is the reduced 
Planck mass \cite{Witten}. 
Modified gravitational theories like $f(R)$ gravity \cite{Staro} 
can also generate a similar interaction between $R$ and 
a gravitational scalar \cite{fRreview}.

In the presence of couplings like $F(\phi)R$, the scalar field 
mediates an extra force with the matter sector \cite{Damour,Fujiibook}. 
We require that such a fifth force is suppressed inside 
the Solar System for compatibility with local gravity 
experiments \cite{Will}. 
The chameleon mechanism \cite{chameleon} 
is one of the means of suppressing the fifth force in regions of the 
high density. The success of this mechanism requires the large field mass 
in local regions of the Universe. 
If the same field is responsible for 
the present cosmic acceleration, its mass needs to be very small
on cosmological scales ($m_{\phi} \sim 10^{-33}$ eV). 
These conditions put strong restrictions on the viable form 
of the scalar potential \cite{Yoko,Mota}. 
Since the field mass $m_{\phi}$ becomes larger in the past,  
we need to fine-tune initial conditions of scalar-field
perturbations in such a way that a heavy oscillating mode 
induced by the large mass is suppressed relative 
to a matter-induced mode \cite{fR1,fR2}.

The Vainshtein mechanism \cite{Vain} is another way for suppressing 
the fifth force around a matter source, which is attributed to 
nonlinear scalar-field self-interactions like $X \square \phi$ 
(where $X$ is the field kinetic energy). 
Such scalar self-interactions arise in the Dvali-Gabadadze-Porrati (DGP) 
model \cite{DGP} as a result of the mixture of a longitudinal graviton $\phi$
(brane-bending mode) and a transverse graviton.
The Vainshtein mechanism is at work for a massless scalar, 
so dark energy models constructed in this vein are usually 
free from the existence of heavy oscillating modes 
in the early cosmological epoch.

In the Minkowski spacetime the field equation of motion following from 
the Lagrangian $L_3=X \square \phi$ is invariant under the Galilean shift 
$\partial_{\mu} \phi \to \partial_{\mu} \phi+b_{\mu}$.
There are other four Lagrangians respecting the Galilean 
symmetry, which contain the linear potential $L_1=\phi$ and 
the kinetic term $L_2=X$ \cite{Nicolis}. On the curved background this 
symmetry is generally broken, but we can derive five 
``covariant Galileon'' Lagrangians with second-order equations 
of motion recovering the Galilean symmetry in the limit 
of Minkowski space-time \cite{Galileons}. 
This is achieved by introducing 
nonminimal  field derivative couplings with curvature 
quantities such as $X^2 R$ and 
$X^2 G_{\mu \nu}\nabla^{\mu}\nabla^{\nu}\phi$, 
where $G_{\mu \nu}$ is the Einstein tensor. 
The generalization of covariant Galileons led to the rediscovery 
of most general scalar-tensor theories with second-order 
equations of motion \cite{Deffayet}-- Horndeski theories \cite{Horndeski}.  

The cosmology for the covariant Galileon without the field 
potential $V(\phi)$ was studied in Refs.~\cite{GS10,DT10}. 
Under two constraints among coefficients of four 
Galileon Lagrangians there exists a tracker solution with a de Sitter 
attractor \cite{DT10}, along which the field derivative $\dot{\phi}$ with 
respect to the time $t$ is 
inversely proportional to the Hubble parameter $H$. 
However the field equation of state for the tracker 
is away from $-1$ during the matter era ($w_{\phi}=-2$), 
so only the late-time tracking solutions can be compatible 
with observational data \cite{Nesseris}.
Even for late-time tracking solutions, the covariant Galileon 
with de Sitter attractors is in tension with 
redshift-space distortion data \cite{Totani} due to the large 
growth rate of matter perturbations \cite{FKT10}.
This tension can be slightly relaxed without assuming the existence 
of late-time de Sitter solutions \cite{Barreira}.

In this paper, we propose a viable modified gravity model of dark energy
compatible with both cosmological and local gravity constraints
by generalizing the covariant Galileon studied in the literature.
We perform the following generalizations: 
(1) the linear potential $V(\phi)=a_1\phi$ is considered in the analysis, 
(2) the nonminimal coupling $F(\phi)R$ and the kinetic term 
$-(1-6q^2)F(\phi)X/2$ with $F(\phi)=e^{-2q\phi/M_{\rm pl}}$ 
are taken into account,
(3) nonlinear field self-interactions beyond the realm of 
Horndeski theories are considered.

The generalization (1) implies that the late-time acceleration of 
the Universe is driven by the potential of a massless scalar 
rather than the time derivative $\dot{\phi}$, so we do not 
need to resort to late-time tracking solutions with restricted 
initial conditions for realizing the cosmic acceleration. 
The cosmology with a light mass Galileon was also studied 
in the presence of the term $L_3=X \square \phi$ \cite{Ali}, 
but our model involves more general nonlinear scalar 
self-interactions presented later.

The generalization (2) allows one for encompassing 
Brans-Dicke (BD) theory \cite{Brans} and dilaton gravity \cite{darkco}
as specific cases. 
The nonminimal coupling $e^{-2q\phi/M_{\rm pl}}R$ induces 
an interaction between the scalar $\phi$ and matter, but 
nonlinear field self-interactions can effectively 
reduce such an interaction.
A similar situation also arises in the DGP model, where 
reduction of the 5-dimensional bulk action to 
the 4-dimensional brane gives rise to an interaction between the 
brane-bending mode $\phi$ and matter as well as the field 
self-interaction $L_3=X \square \phi$ \cite{Luty}. 
In fact, the nonminimal coupling $e^{-2q\phi/M_{\rm pl}}R$ 
in the presence of the Galileon term $L_3$
can reproduce many features of the DGP model \cite{Chow}.

For the generalization (3) we introduce nonlinear field self-interactions in the 
framework of Gleyzes-Langlois-Piazza-Vernizzi (GLPV) \cite{GLPV} theories 
whose deviation from Horndeski theories is weighed by a parameter $\alpha_{\rm H}$.
Extension of Horndeski theories to GLPV theories on the flat 
Friedmann-Lema\^{i}tre-Robertson-Walker (FLRW) background 
does not increase the number of propagating scalar degrees of freedom \cite{GLPV,Hami}.
GLPV theories exhibit several interesting properties such as the mixing of scalar/matter 
propagation speeds \cite{Gergely,KT14} and the appearance of 
heavy oscillating modes of gravitational potentials \cite{Koyama,Tsuji15}. 
Hence it is of interest to study in detail whether our model is cosmologically 
viable without the problems of instabilities and fine-tuned initial conditions.

Recently, it was found that the model approaching a nonzero $\alpha_{\rm H}$ 
at the center of a spherically symmetric body ($r=0$) is plagued by the 
appearance of a conical singularity \cite{DKT15}. 
Our model satisfies the condition 
$\alpha_{\rm H} (r \to 0)=0$, so the conical singularity 
is absent. The Vainshtein mechanism can be at work in our 
model, but it is not clear whether field self-interactions 
compatible with cosmological constraints lead to 
the recovery of GR in the Solar System.
In this paper we elucidate a viable parameter space in which 
the model is consistent with both cosmological and local 
gravity constraints.

This paper is organized as follows.
In Sec.~\ref{modelsec} we present the Lagrangian of our model,  
which belongs to a class of GLPV theories without a conical singularity.
In Sec.~\ref{backsec}  the background cosmology is discussed 
by paying attention to the evolution of field self-interacting terms.
In Sec.~\ref{cosmosec} we study the stability of perturbations
by deriving the scalar propagation speed and then 
investigate the evolution of gravitational 
potentials and matter perturbations.
In Sec.~\ref{vasec} we identify a viable parameter space for 
the amplitude of nonlinear field self-interactions by discussing the 
Vainshtein screening in the Solar System.
Sec.~\ref{consec} is devoted to conclusions.

%============================================
%section II
\section{Model}
\label{modelsec} 
%===========================================

Our model belongs to a subclass of GLPV theories \cite{GLPV} 
given by a sum of the following three Lagrangians:
\ba
L_{2} &=&A_{2}(\phi ,X)\,,  
\label{L2} \\
L_{3} &=&\left( C_{3}+2XC_{3,X} \right)
\square \phi +XC_{3,\phi}\,,  
\label{L3} \\
L_{4} &=&B_{4} R-\frac{A_{4}+B_{4}}
{X}\left[ (\square \phi )^{2}-\nabla ^{\mu} 
\nabla ^{\nu}\phi \nabla_{\mu}\nabla_{\nu}\phi \right]  \nonumber \\
&&+\frac{2\left(A_{4}+ B_{4}-2XB_{4,X} \right)}
{X^{2}} ( \nabla ^{\mu}\phi \nabla ^{\nu}\phi 
\nabla _{\mu}\nabla _{\nu}\phi\,
\square \phi \nonumber \\
& &-\nabla ^{\mu}\phi \nabla _{\mu}\nabla _{\nu}\phi 
\nabla _{\sigma}\phi
\nabla ^{\nu}\nabla ^{\sigma}\phi )\,,
\label{L4} 
\ea
with $\nabla_{\mu}$ is the covariant derivative, 
$A_2, C_3, A_4, B_4$ are functions of 
the scalar field $\phi$ and its kinetic energy 
$X=\nabla_{\mu}\phi \nabla^{\mu}\phi$, and 
$B_{4,X}=\partial B_4/\partial X$.
If the condition $A_{4}+ B_{4}=2XB_{4,X}$ is satisfied, 
the Lagrangian (\ref{L4}) belongs to a subclass  
of Horndeski theories. To quantify the departure from 
Horndeski theories, we introduce the parameter
\be
\alpha_{\rm H} \equiv \frac{2XB_{4,X}-A_4-B_4}{A_4}\,.
\label{alH}
\ee
In full GLPV theories there exists an additional Lagrangian $L_5$ to 
Eqs.~(\ref{L2})-(\ref{L4}), but we do not take into account such a 
contribution in this paper. This is not only for simplicity but it is 
attributed to the fact that the Lagrangian $L_5$ tends to 
prevent the success of the Vainshtein mechanism \cite{Kimura,Kase13}.

Choosing the unitary gauge on the FLRW background in which the scalar field $\phi$ depends on the time $t$ alone, the 
Lagrangian $L$ can be expressed in terms of geometric scalars 
appearing in the 3+1 Arnowitt-Deser-Misner (ADM) decomposition 
of space-time. The ADM formalism is based on the line element
$ds^2=g_{\mu \nu}dx^{\mu}dx^{\nu}= -N^{2}dt^{2}
+h_{ij}(dx^{i}+N^{i}dt)(dx^{j}+N^{j}dt)$, 
where $N$ is the lapse, $N^i$ is the shift, and 
$h_{ij}$ is the 3-dimensional spatial metric \cite{ADM}.
We define the extrinsic curvature and the intrinsic curvature, 
respectively, as $K_{\mu \nu}=h^{\lambda}_{\mu} n_{\nu;\lambda}$ 
and ${\cal R}_{\mu \nu}={}^{(3)}R_{\mu \nu}$, where
$n_{\mu}=(-N,0,0,0)$ is a normal vector orthogonal 
to constant time hyper-surfaces $\Sigma_t$ and 
${}^{(3)}R_{\mu \nu}$ is the 3-dimensional Ricci tensor on $\Sigma_t$.
In terms of the scalar quantities $K=g^{\mu \nu}K_{\mu \nu}$, 
${\cal S}=K^{\mu \nu}K_{\mu \nu}$, and ${\cal R}=g^{\mu \nu}{\cal R}_{\mu \nu}$, 
the Lagrangian $L=L_2+L_3+L_4$ can be expressed 
in the form \cite{building,GLPV,Tsujilec}
\be
L=A_2(N,t)+A_3(N,t)K+A_4(N,t)(K^2-{\cal S})
+B_4(N,t){\cal R}\,,
\label{Lag}
\ee
where  
\be
A_3=2|X|^{3/2} \left(C_{3,X}+\frac{B_{4,\phi}}{X} 
\right)\,.
\label{A3}
\ee
On the FLRW background, we have $X<0$ and hence 
$A_3=2(-X)^{3/2}C_{3,X}-2\sqrt{-X}B_{4,\phi}$.
On the spherically symmetric and static background ($X>0$),
it follows that $A_3=2X^{3/2}C_{3,X}+2\sqrt{X}
B_{4,\phi}$ \cite{DKT15}.

Taking into account matter minimally coupled to gravity 
(described by the Lagrangian $L_m$), the action 
in the unitary gauge on the flat FLRW background 
can be expressed as
\be
S=\int d^4 x \sqrt{-g}\,L (N,K,{\cal S},{\cal R};t)
+\int d^4 x \sqrt{-g}\,L_m\,,
\label{act}
\ee
where $g$ is the determinant of $g_{\mu \nu}$ and 
$L$ is given by Eq.~(\ref{Lag}). 
In the following we focus on the model described by 
\ba
A_2 &=& -\frac12 \omega(\phi)X-V(\phi)\,,\label{A2} \\
C_3 &=& a_3 X\,,\\
A_4 &=&-\frac12 M_{\rm pl}^2 F(\phi)+a_4X^2\,,\label{A4}\\
B_4 &=& \frac12 M_{\rm pl}^2 F(\phi)+b_4X^2\,,
\label{B4}
\ea
where $a_3$, $a_4$, $b_4$ are constants, 
$\omega(\phi)$, $V(\phi)$, and $F(\phi)$ are 
functions of $\phi$,
and $M_{\rm pl}$ is the reduced Planck mass. 
The function $A_3$ reads
\be
A_3=2|X|^{3/2} \left(a_3+\frac{M_{\rm pl}^2F_{,\phi}}{2X} 
\right)\,. \label{A3d}
\ee

The Lagrangian $L$ with the functions (\ref{A2})-(\ref{B4}) 
can accommodate several theories known in the literature. 
BD theory \cite{Brans} corresponds to 
the functions $\omega(\phi)=(1-6q^2)F(\phi)$, 
$F(\phi)=e^{-2q\phi/M_{\rm pl}}$, $V(\phi)=0$, 
$a_3=0$, $a_4=0$, $b_4=0$ with the BD parameter 
related to a constant $q$ as 
$\omega_{\rm BD}=(1-6q^2)/(4q^2)$ \cite{chameleon,Yoko}. 
If we transform the action of BD theory to that in the Einstein 
frame, the constant $q$ plays the role of a coupling between 
the field $\phi$ and nonrelativistic matter.
Dilaton gravity \cite{Witten} is the special case of BD theory with 
$\omega_{\rm BD}=-1$, i.e., $q^2=1/2$. 

For a {\it massless} BD field the BD parameter is constrained 
to be $\omega_{\rm BD}>4 \times 10^4$ from solar-system 
constraints \cite{Will}, which translates to the condition 
$|q|<2.5 \times 10^{-3}$.
In the case of BD theory with $|q| \gtrsim 10^{-3}$, 
we require either a massive scalar 
potential or nonlinear field self-interactions for  
the compatibility with local gravity experiments.
Due to the problem of initial conditions associated with a heavy 
oscillating mode for the massive scalar, we resort to the latter nonlinear
field self-interactions for suppressing the fifth force inside 
the Solar System. 
In the following we shall focus on the coupling
\be
|q| \gtrsim 10^{-3}\,,
\ee
in which case the Vainshtein screening is required for 
the compatibility with local gravity experiments.

The $X^2$ terms in Eqs.~(\ref{A4}) and (\ref{B4}), which arise 
for the covariant Galileon \cite{Galileons}, describe such field 
self-interactions. The covariant Galileon belongs to a sub-class 
of Horndeski theories, in which case there is the particular 
relation $a_4=3b_4$ from the condition $\alpha_{\rm H}=0$.
In GLPV theories this relation no longer holds, but 
it was found that the Vainshtein mechanism can operate to 
suppress the fifth force even for $a_4 \neq 3b_4$ \cite{DKT15}. 
Meanwhile, even the small departure from Horndeski theories 
potentially gives rise to a large deviation from it at the level 
of cosmological perturbations \cite{KT14,Koyama,Tsuji15}, 
so there should be some cosmological bound 
on the parameter $\alpha_{\rm H}$. 
In this paper we shall study a parameter space in which 
the model is consistent with both local gravity and 
cosmological constraints.

Since we would like to focus on a massless scalar field, we 
adopt the linear potential $V(\phi)=a_1 \phi$ as it appears 
for the Minkowski Galileon \cite{Nicolis}. 
Provided that the initial displacement 
of the field is in the range $|\phi| \gtrsim M_{\rm pl}$, 
this potential can drive the temporal cosmic acceleration. 
In Ref.~\cite{DT10} the Galileon cosmology was studied 
for $V(\phi)=0$ and $q=0$ in the presence of an additional 
Lagrangian $L_5$. In this case, at the expense of imposing 
two constraints among coefficients of each Galileon Lagrangian, 
it is possible to realize de Sitter solutions with $\dot{\phi}={\rm constant}$.
The difference from this approach is that we do not impose such constraints
and hence the late-time cosmic acceleration is driven by 
the linear potential $V(\phi)=a_1 \phi$, 
but not by the kinetic term $\dot{\phi}$.

In summary, we focus on the theories given by 
Eqs.~(\ref{A2})-(\ref{B4}), with the functions
\ba
\omega(\phi)
&=& (1-6q^2)F(\phi)\,,\\
F(\phi)
&=&e^{-2q\phi/M_{\rm pl}}\,, \\
V(\phi)
&=&a_1\phi\,,
\label{linear}
\ea
without necessarily imposing the relation $a_4=3b_4$. 
Our model is not plagued by the problem of a conical 
singularity at the center of a compact object because 
it satisfies the condition $\alpha_{\rm H}(r \to 0)=0$ \cite{DKT15}. 
This is not the case for the model studied in Ref.~\cite{Koyama} 
where $\alpha_{\rm H}$ is a non-zero constant at $r=0$.

%===========================================
%section III
\section{Background cosmology}
\label{backsec} 
%===========================================

On the flat FLRW background described by the line element 
$ds^2=-dt^2+a^2(t)\delta_{ij}dx^idx^j$, 
the background equations of motion 
can be derived by varying the action (\ref{act}) 
with respect to $N$ and $a$. 
Taking into account a perfect-fluid matter with energy density 
$\rho$ and pressure $P$, the resulting equations of motion are 
given by $\bar{L}+L_{,N}-3H{\cal F}=\rho$ and 
$\bar{L}-\dot{\cal F}-3H{\cal F}=-P$ \cite{building}, where 
$H \equiv \dot{a}/a$ is the Hubble parameter, 
${\cal F} \equiv L_{,K}+2HL_{,{\cal S}}$,  and 
a bar represents quantities of the background.
For the theories given by the functions
(\ref{A2})-(\ref{A3d}) they reduce, respectively, to  
\ba
& & 3M_{\rm pl}^2 \left( H^2 F+H\dot{\phi} F_{,\phi} 
\right) \nonumber \\
& &=\frac{\omega}{2}\dot{\phi}^2+V
+18a_3H \dot{\phi}^3+30a_4H^2\dot{\phi}^4+\rho\,,\label{back1}\\
& & M_{\rm pl}^2 \left(3H^2F+2\dot{H}F+2HF_{,\phi}
\dot{\phi}+F_{,\phi} \ddot{\phi}+F_{,\phi \phi} \dot{\phi}^2 \right)  
\nonumber  \\
& &= -\frac12 \omega \dot{\phi}^2+V+
6a_3\dot{\phi}^2 \ddot{\phi} \nonumber \\
& &\quad+2a_4 \dot{\phi}^3 
\left[ 8H \ddot{\phi}+(3H^2+2\dot{H})\dot{\phi} \right]-P\,.
\label{back2}
\ea
In deriving Eqs.~(\ref{back1}) and (\ref{back2}), we have assumed 
that the field derivative is in the range $\dot{\phi}>0$.
The matter sector obeys the continuity equation 
\be
\dot{\rho}+3H(\rho+P)=0\,.
\ee
\subsection{Dynamical equations of motion}

We study the background cosmology in the presence of 
nonrelativistic matter (density $\rho_m$ and
pressure $P_m=0$) and radiation (density $\rho_r$ and
pressure $P_r=\rho_r/3$). In doing so, it is convenient to 
introduce the dimensionless variables
\ba
&&
x_1=\frac{\dot{\phi}}{\sqrt{6}M_{\rm pl}H}\,,\qquad
x_2=\frac{V}{3M_{\rm pl}^2 H^2F}\,,\nonumber \\
&&
x_3=\frac{6a_3 \dot{\phi}^3}{M_{\rm pl}^2 HF}\,,\qquad
x_4=\frac{10a_4\dot{\phi}^4}{M_{\rm pl}^2F}\,,\nonumber \\
& &
x_5=\frac{\sqrt{\rho_r}}{\sqrt{3F}M_{\rm pl}H}\,,\qquad
\lambda=-\frac{M_{\rm pl}V_{,\phi}}{V}\,.
\label{varidef}
\ea

Defining $\Omega_m=\rho_m/(3FM_{\rm pl}^2H^2)$ 
and $\Omega_r=x_5^2$, we can rewrite 
Eq.~(\ref{back1}) as
\be
\Omega_m=1-(1-6q^2)x_1^2-2\sqrt{6}qx_1-x_2-x_3-x_4
-\Omega_r\,.
\ee
The variables defined in Eq.~(\ref{varidef}) obey  
\ba
x_1' &=& x_1 \left( \epsilon_{\phi}
-h \right)\,,\label{x1eq}\\
x_2' &=& x_2 \left[ \sqrt{6} (2q-\lambda) x_1 
-2h \right]\,,\\
x_3' &=& x_3 \left( 2\sqrt{6}qx_1
+3\epsilon_{\phi}-h \right)\,,\label{x3eq}\\
x_4' &=& x_4 \left( 2\sqrt{6}qx_1
+4\epsilon_{\phi} \right)\,,\label{x4eq}\\
x_5' &=& x_5  \left( \sqrt{6}qx_1
-2-h \right)\,,\label{x5eq}\\
\lambda' &=& 
\sqrt{6} \lambda^2 x_1\,,
\ea
where a prime represents a derivative with 
respect to ${\cal N}=\ln a$, and 
\be
\epsilon_{\phi} \equiv \frac{\ddot{\phi}}{H \dot{\phi}}\,,
\qquad h \equiv \frac{\dot{H}}{H^2}\,.
\ee
Taking the time derivative of Eq.~(\ref{back1}) and solving 
for $\ddot{\phi}$ and $\dot{H}$, we obtain 
\begin{widetext}
\ba
\epsilon_{\phi}
&=&
[3\sqrt{6}x_1^2 \{ 5x_3-20+2q^2 (20+5x_3-4x_4)+12x_4 \}
+12q x_1^3 \{ 25-2x_4+6q^2(2x_4-5 )\}+\sqrt{6} \{ -24x_2x_4\nonumber \\
& &+3x_3 (x_4-5x_2-5)
+(5x_3+8x_4)x_5^2 \} 
-12x_1 \{ \lambda x_2(x_4-5)+5q(1+3x_2+2x_3
+3x_4-x_5^2)\}]/(\sqrt{6}{\cal D}),\label{ddotphi}\\
h
&=& -[30(1-8q^2+12q^4)x_1^4+15x_3^2+2\sqrt{6}q
(6q^2-1)x_1^3 (5x_3+8x_4-10) 
+\sqrt{6}x_1 \{ 5x_3(2q-\lambda x_2)
+8x_4(q-\lambda x_2) \}\nonumber \\
& &+10x_3 (3-3x_2+3x_4+x_5^2)
+12x_4(3-3x_2+x_4+x_5^2)
+2x_1^2 (15-15x_2+30x_3+39x_4+5x_5^2) \nonumber \\
& &-60q x_1^2 \{ \lambda x_2-q(1+3x_2-2x_3-3x_4-x_5^2) \}]
/{\cal D}\,,\label{dotH} \\
{\cal D}
&=&
5x_3(4+x_3)+12(2+x_3)x_4+8x_4^2+4\sqrt{6}qx_1 
(5x_3+8x_4)+4x_1^2 \{5+(6q^2-1)x_4 \}\,.
\ea
\end{widetext}

If $\lambda$ is a constant, i.e., for the exponential 
potential $V(\phi)=V_0e^{-\lambda \phi/M_{\rm pl}}$, 
the fixed points can be derived by setting 
$x_i'=0$ in Eqs.~(\ref{x1eq})-(\ref{x5eq}).
For the linear potential (\ref{linear}), 
the parameter $\lambda$ reads 
\be
\lambda=-\frac{M_{\rm pl}}{\phi}\,.
\label{lamlinear}
\ee
We focus on the case in which 
the field evolves along the potential (\ref{linear}) 
with $a_1<0$. Provided $\phi<0$, we have that  
$V(\phi)>0$ and $\lambda>0$. 
As $\phi$ approaches 0 from $\phi<0$, 
the variable $\lambda$ continues to grow. 
The late-time cosmic acceleration can be realized for 
$\lambda \lesssim 1$, so the initial field displacement 
needs to be in the range $|\phi| \gtrsim M_{\rm pl}$.

\subsection{Instantaneous fixed points for $a_3=a_4=0$}

As long as the variation of $\phi$ is not large such that 
$\dot{\lambda}/(H\lambda) \lesssim 1$ (which translates 
to the condition $\sqrt{6}\lambda x_1 \lesssim 1$),
the fixed points of the system (\ref{x1eq})-(\ref{x5eq}) 
can be regarded as instantaneous ones \cite{Nelson}.
For the theories with $a_3=a_4=0$ 
and $\lambda \lesssim 1$, there exist the following 
instantaneous fixed points relevant to 
(a) radiation-dominated, (b) matter-dominated, and 
(c) scalar-field-dominated epochs.
\ba
\hspace{-1.0cm}
& &
{\rm (a)}~~(x_1, x_2, x_5)=(0,0,1)\,,\\
\hspace{-1.0cm}
& &
{\rm (b)}~~(x_1, x_2, x_5)=\left( 
-\frac{\sqrt{6}q}{3(1-2q^2)},0,0 
\right)\,,\label{pMDE}\\
\hspace{-1.0cm}
& &
{\rm (c)}~~(x_1, x_2, x_5) \nonumber \\
\hspace{-1.0cm}
& &~~~~~=\left( 
\frac{\sqrt{6}(4q-\lambda)}{6(4q^2-\lambda q-1)},
\frac{6-\lambda^2+8q\lambda-16q^2}
{6(4q^2-q\lambda-1)^2},0 \right).
\ea
The parameters $\Omega_m$, $\Omega_r$, 
$\Omega_{\rm DE}=1-\Omega_m-\Omega_r$, 
and the effective equation of state 
$w_{\rm eff}=-1-2\dot{H}/(3H^2)$ for each point are 
given, respectively, by 
\ba
\hspace{-1.0cm}
& &
{\rm (a)}~~\Omega_m=0,\quad \Omega_r=1,\quad 
\Omega_{\rm DE}=0,\quad
w_{\rm eff}=\frac13\,,\\
\hspace{-1.0cm}
& &
{\rm (b)}~~\Omega_m=\frac{3-2q^2}{3(1-2q^2)^2},
\quad \Omega_r=0,
\nonumber \\
\hspace{-1.0cm}
& &~~~~~~
\Omega_{\rm DE}=\frac{2q^2 (6q^2-5)}{3(1-2q^2)^2}\,,
\quad
w_{\rm eff}=\frac{4q^2}{3(1-2q^2)}\,,\label{pMDE2}\\
\hspace{-1.0cm}
& &
{\rm (c)}~~\Omega_m=0,
\quad \Omega_r=0,\quad \Omega_{\rm DE}=1\,, \nonumber \\
\hspace{-1.0cm}
& & ~~~~~~
w_{\rm eff}=-\frac{20q^2-9q\lambda-3+\lambda^2}
{3(4q^2-q\lambda-1)}\,.
\ea

The point (b) corresponds to the $\phi$-matter-dominated-epoch 
($\phi$MDE) \cite{Luca} in the Jordan frame \cite{Yoko}. 
We require the condition $|q| \ll 1$ to mimic the standard 
matter era ($\Omega_m \simeq 1$ and $w_{\rm eff} \simeq 0$).
For a positive field kinetic energy ($\dot{\phi}>0$), 
the $\phi$MDE is present if 
\be
q<0\,.
\ee
We shall focus on this case throughout this paper.
In the context of coupled dark energy, the model with negative 
$q$ of the order of $-0.01$ is even favored from the Planck 
data \cite{Valeria} (the opposite notation of $q$ is 
used in Ref.~\cite{Valeria}).

The point (c) can lead to the cosmic acceleration 
($w_{\rm eff}<-1/3$) for 
\be
-\sqrt{2}+4q<\lambda<\sqrt{2}+4q\,.
\label{accon}
\ee
Provided that the parameter $\lambda=-M_{\rm pl}/\phi$ satisfies 
the condition (\ref{accon}), the Universe enters the 
accelerated epoch. 
As $\phi$ approaches 0 the condition (\ref{accon}) starts 
to be violated, so the cosmic acceleration ends at some point. 
In summary, for the theories with $a_3=a_4=0$, 
we have the cosmological sequence of 
instantaneous fixed points (a)$\to$(b)$\to$(c) 
with a final exit from the accelerating regime.

\subsection{Cosmological dynamics for 
$a_3 \neq 0, a_4 \neq 0$}

Let us discuss the cosmological dynamics for the theories 
with $a_3 \neq 0$ and $a_4 \neq 0$.
We begin with the study about the behavior of terms 
$x_3$ and $x_4$ relative to the standard normalized
kinetic term $x_1^2$.
During the radiation and the deep matter eras, 
we consider the situation in which the parameters 
$x_1^2,x_2,x_3,x_4$ are much smaller than 1 
with $|q|<O(1)$.
{}From Eqs.~(\ref{ddotphi}) and (\ref{dotH})
it then follows that 
\ba
\hspace{-0.4cm}
& &
\frac{\ddot{\phi}}{H\dot{\phi}}
\simeq \frac{-15(4x_1^2+x_3)+(5x_3+8x_4)x_5^2+\beta}
{4(5x_1^2+5x_3+6x_4)},\label{ddotphi2} \\
\hspace{-0.4cm}
& & 
\frac{\dot{H}}{H^2} \simeq -\frac32-\frac12 x_5^2\,,
\label{dotH2}
\ea
where $\beta=-10\sqrt{6}qx_1(1-x_5^2)$.
The term $\beta$ gives rise to a positive contribution 
to $\ddot{\phi}/(H\dot{\phi})$ for $x_1>0$ and $x_5^2<1$.
Under the conditions $x_1^2 \gg \{ |x_3|, |x_4|\}$ we have 
$\ddot{\phi}/(H\dot{\phi}) \simeq -3-\sqrt{6}
q(1-x_5^2)/(2x_1)$. Substituting this  
relation and Eq.~(\ref{dotH2}) with $x_5 \simeq 0$ 
into Eq.~(\ref{x1eq}), it follows that there exists the 
$\phi$MDE characterized by $x_1 \simeq -\sqrt{6}q/3$. 
This shows that, for $x_1^2$ greater than $|x_3|$ and $|x_4|$,   
the presence of coupling $q$ leads to the approach to 
the $\phi$MDE. Since $x_3/x_1^2 \propto H\dot{\phi}/F 
\propto H^2/F$ and $x_4/x_1^2 \propto (H\dot{\phi})^2/F 
\propto H^4/F$ during the $\phi$MDE, both $x_3$ and $x_4$ 
decrease rapidly relative to $x_1^2$. 
Hence the cosmological dynamics is similar to that for 
$a_3=a_4=0$.

\begin{figure}
\includegraphics[height=3.2in,width=3.2in]{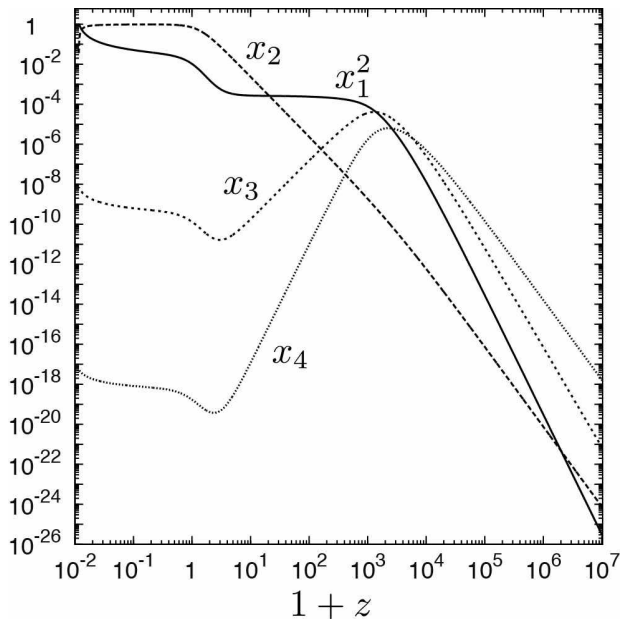}
\includegraphics[height=3.2in,width=3.4in]{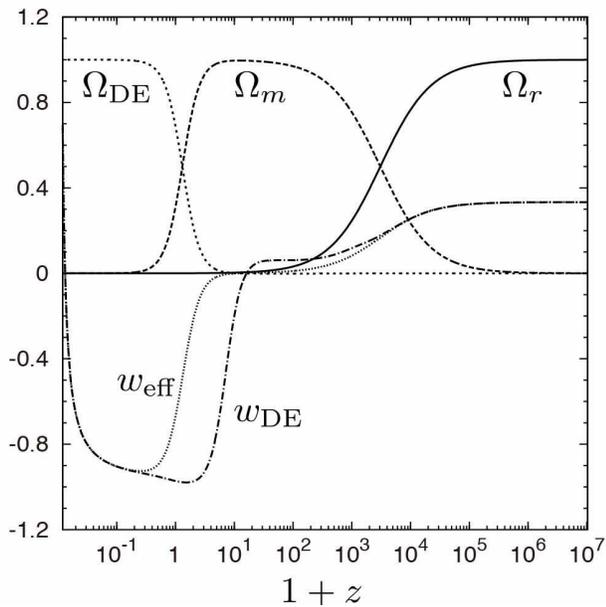}
\caption{\label{xifig}
Evolution of $x_1^2, x_2, x_3, x_4$ (top) and 
$\Omega_r$, $\Omega_m$, $\Omega_{\rm DE}$, 
$w_{\rm eff}$, $w_{\rm DE}$ (bottom) versus 
$1+z$ for $q=-0.02$ with the initial conditions 
$x_1=1.66 \times 10^{-13}$, $x_2=7.02 \times 10^{-25}$,
$x_3=6.51 \times 10^{-22}$, $x_4=1.54 \times 10^{-18}$, 
$x_5=0.99985$, and $\lambda=0.3$ at the redshift 
$z=1.02 \times 10^7$. 
The present epoch ($z=0$) is identified 
by $\Omega_{\rm DE}=0.68$.}
\end{figure}

Let us consider the case in which the terms $|x_3|$ and 
$|x_4|$ are larger than the order of $x_1^2$ during the 
radiation era ($x_5 \simeq 1$ and $\dot{H}/H^2 \simeq -2$).
If $x_3$ is the dominant contribution to Eq.~(\ref{ddotphi2}), 
it follows that $\ddot{\phi}/(H\dot{\phi}) \simeq -1/2$. 
Integrating Eq.~(\ref{x1eq}), (\ref{x3eq}),
and (\ref{x4eq}), we obtain the solutions $x_1 \propto e^{3{\cal N}/2}$, 
$x_3 \propto e^{{\cal N}/2}$, and
$x_4 \propto e^{-2{\cal N}}$, respectively.
If $x_4$ is the dominant term in Eq.~(\ref{ddotphi2}), 
we have that $\ddot{\phi}/(H\dot{\phi}) \simeq 1/3$ and hence
$x_1 \propto e^{7{\cal N}/3}$, 
$x_3 \propto e^{3{\cal N}}$, and
$x_4 \propto e^{4{\cal N}/3}$. 
In both cases, the terms $x_3$ and $x_4$ evolve slowly compared to $x_1^2$. 
In the above estimation we have ignored the contribution from 
$\beta$, but this works to enhance the growth rates of 
$x_1$, $x_3$, and $x_4$ due to the increase of $\ddot{\phi}/(H\dot{\phi})$.
Note that $x_1$ does not grow for the theories 
with $a_3=a_4=0$ and $q=0$, in which case 
$\ddot{\phi}/(H\dot{\phi})=-3$ and $x_1 \propto e^{-{\cal N}}$.  

In Fig.~\ref{xifig} we plot the evolution of 
$x_1^2, x_2, x_3, x_4$ for $q=-0.02$ with the initial conditions 
satisfying $x_4 \gg x_3 \gg x_1^2$. 
As estimated above, the growth rate of $x_1^2$ 
during the radiation-dominated epoch
is larger than those of $x_3$ and $x_4$. 
The variable $x_3$ also increases faster than $x_4$.
For the initial conditions employed in Fig.~\ref{xifig}, 
$x_1^2$ catches up with $x_3$ and 
$x_4$ by the end of the radiation era, which is followed 
by the $\phi$MDE characterized by $x_1^2 \simeq 2q^2/3$. 
As we already mentioned, the ratios $x_3/x_1^2$ and 
$x_4/x_1^2$ during the $\phi$MDE are proportional 
to $H^2$ and $H^4$ for nearly constant $F$, respectively, 
so that they decrease as $x_3/x_1^2 \propto t^{-2}$ and 
$x_4/x_1^2 \propto t^{-4}$.
This property can be clearly seen in Fig.~\ref{xifig}.

While we have chosen the coupling $q=-0.02$ in 
Fig.~\ref{xifig}, the larger values of $|q|$ like $|q| > O(0.1)$ 
give rise to a stronger modification to the distance to the last scattering 
surface of Cosmic Microwave Background (CMB).
This is analogous to what happens in the coupled dark energy scenario 
\cite{CMB,Valeria}, so the coupling $|q|$ would be smaller than the order of 0.1.
To put precise constraints on $q$, we require a joint analysis based on CMB 
and other data, but this is beyond the scope of our present work.

Even though the potential energy $V(\phi)$ gets larger than 
the kinetic energy $\dot{\phi}^2/2$ (i.e., $x_2>x_1^2$) 
in the numerical simulation of Fig.~\ref{xifig} 
(around the redshift $z \equiv 1/a-1=20$), the cosmic acceleration starts 
to occur  at a later epoch after the potential energy dominates 
over the matter energy density 
(i.e., $x_2 \gtrsim \Omega_m$).
In the bottom panel of Fig.~\ref{xifig} we plot the evolution 
of $w_{\rm eff}$ as well as the density parameters $\Omega_{\rm DE}$, 
$\Omega_m$, and $\Omega_r$. 
The effective equation of state evolves from 
$w_{\rm eff} \simeq 1/3$ (radiation era) to 
$w_{\rm eff} \simeq 0$ (matter era), and then 
it enters the epoch of cosmic acceleration 
($w_{\rm eff}<-1/3$) around $z=0.6$. 
Since $H$ does not vary much after the onset of 
cosmic acceleration, the ratios $x_3/x_1^2$ 
and $x_4/x_1^2$ stay nearly constant in this regime.

\begin{figure}
\includegraphics[height=3.2in,width=3.3in]{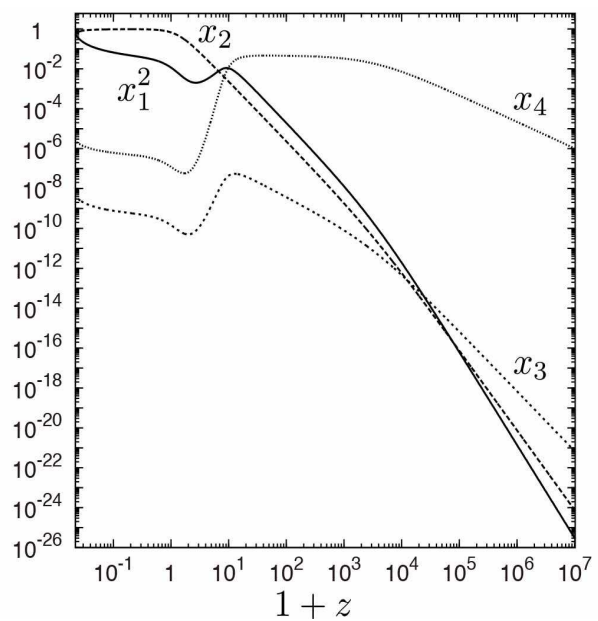}
\includegraphics[height=3.2in,width=3.25in]{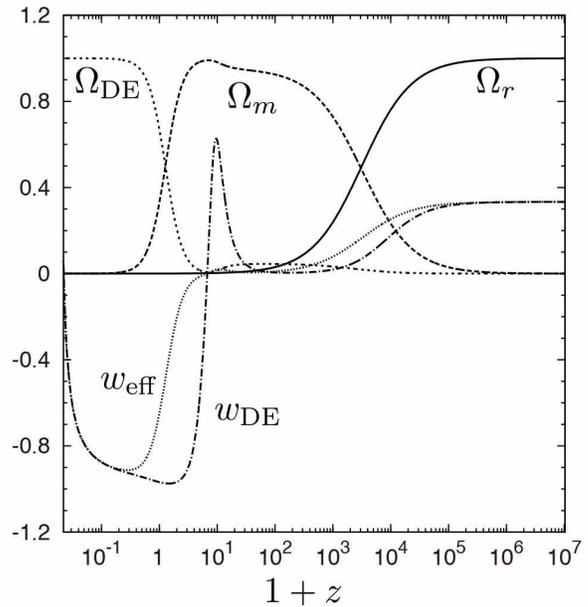}
\caption{\label{xifig2}
The same as Fig.~\ref{xifig}, but the different choice of 
$a_4$ such that $x_4=1.0 \times 10^{-6}$ at the redshift 
$z=1.03 \times 10^{7}$.}
\end{figure}

We introduce the energy density $\rho_{\rm DE}$ and 
the pressure $P_{\rm DE}$ of dark energy by rewriting 
Eqs.~(\ref{back1}) and (\ref{back2}), as
\ba
& & 3M_{\rm pl}^2H^2F_0=\rho_{\rm DE}+\rho\,,\\
& & M_{\rm pl}^2 \left( 3H^2 F_0+2\dot{H} F_0 \right)
=-P_{\rm DE}-P\,,
\ea
where $F_0$ is the present value of $F$. 
Note that $\rho_{\rm DE}$ and $P_{\rm DE}$
obey the standard continuity equation 
$\dot{\rho}_{\rm DE}+3H(\rho_{\rm DE}+P_{\rm DE})=0$.
The dark energy equation of state 
$w_{\rm DE}=P_{\rm DE}/\rho_{\rm DE}$ is 
related to $w_{\rm eff}$, as
\be
w_{\rm DE}=\frac{w_{\rm eff}-(\Omega_r/3)(F/F_0)}
{1-(\Omega_m+\Omega_r) (F/F_0)}\,.
\label{wDEe}
\ee
During the radiation-dominated epoch 
($w_{\rm eff}=1/3$, $\Omega_r=1$, $\Omega_m=0$), 
this simply reduces to $w_{\rm DE}=1/3$. 
For the $\phi$MDE characterized by Eq.~(\ref{pMDE2}), 
it follows that 
\be
w_{\rm DE}=\frac{4q^2(1-2q^2)}
{3(1-F/F_0)-2q^2(6-6q^2-F/F_0)}\,,
\label{wdephi}
\ee
where $F/F_0$ is smaller than 1 (because $F$ increases in time).
If the term $3(1-F/F_0)$ is the dominant contribution to 
the denominator of Eq.~(\ref{wdephi}), it follows that 
$w_{\rm DE} \simeq w_{\rm eff}/(1-F/F_0)>w_{\rm eff} \simeq 4q^2/3$. 
In fact, the numerical simulation of Fig.~\ref{xifig} shows that 
$w_{\rm DE}>w_{\rm eff}$ during the $\phi$MDE. 
After $x_2$ gets larger than $x_1^2$ around $z=20$, 
$w_{\rm DE}$ starts to decrease to negative values.
In Fig.~\ref{xifig} the decrease of $w_{\rm DE}$ occurs earlier than that of $w_{\rm eff}$. 

Since the field evolves along the linear potential, 
the Universe eventually enters the epoch in which  
the parameter $\lambda$ does not satisfy
the condition (\ref{accon}).
In Fig.~\ref{xifig} we find that both $w_{\rm DE}$ 
and $w_{\rm eff}$ begin to increase after 
reaching their minima.
This increase comes from the fact that the field starts to 
evolve faster along the potential. After $\Omega_m$ decreases 
sufficiently, Eq.~(\ref{wDEe}) shows that $w_{\rm DE}$ is 
practically identical to $w_{\rm eff}$.
The cosmic acceleration finally terminates
after $w_{\rm eff}$ crosses the value $-1/3$ again 
in the future. 
For decreasing $|q|$ the value $x_1^2$ corresponding 
to the $\phi$MDE is smaller, so the transition 
from the acceleration to the deceleration occurs 
at a later cosmological epoch.

The numerical simulation of Fig.~\ref{xifig}  
corresponds to the case in which $x_1^2$ catches up with
$x_3,x_4$ around the end of the radiation era.
For larger values of $a_3$ and $a_4$, it happens that 
the terms $x_3$ and $x_4$ dominate over $x_1^2$ 
even in the matter era. 
In Fig.~\ref{xifig2} we show such an example, where
the initial condition of $x_4$ is chosen to be much larger 
than that used in Fig.~\ref{xifig}. 
In this case, $x_4$ dominates over $x_1^2$ by the redshift 
$z \approx 10$, after which $x_4$ and $x_3$ decrease
by the onset of cosmic acceleration.

{}From the lower panel of Fig.~\ref{xifig2} we find that 
the contribution of $x_4$ to $\Omega_{\rm DE}$ is quite 
large even in the early matter era, which leads to 
the suppression of $\Omega_m$. 
Moreover, the dark energy equation of state $w_{\rm DE}$ 
exhibits a peculiar evolution around $x_1^2 \approx x_4$. 
If we choose larger initial values of $x_4$ 
than that used in Fig.~\ref{xifig2}, we do not have 
a proper matter-dominated epoch due to the suppression 
of $\Omega_m$. To avoid this behavior for $|q| \lesssim 0.1$, 
the present value of $x_4$ is constrained to be 
\be
x_4(z=0) \lesssim 10^{-6}\,.
\label{x4con1}
\ee
Strictly speaking, this upper bound tends to decrease 
for smaller $|q|$, but even for $|q| \ll 1$, 
it is at most of the order of $10^{-7}$.
If we consider the case where $x_3$ dominates over 
$x_1^2$ and $x_4$ after the radiation era, the condition
for realizing an appropriate matter-dominated 
epoch is given by 
\be
x_3(z=0) \lesssim 10^{-4}\,.
\label{x3con1}
\ee

The conditions (\ref{x4con1}) and (\ref{x3con1}) should be 
regarded as approximate criteria for the validity of our model 
at the background level.

%============================================
%section III
\section{Dynamics of cosmological perturbations}
\label{cosmosec} 
%===========================================

We proceed to the discussion of cosmological perturbations to put 
further constraints on the parameters of our model.
On the flat FLRW background, we consider four scalar metric 
perturbations $A$, $\psi$, $\zeta$ and tensor perturbations 
$\gamma_{ij}$ given by the line element 
\ba
\hspace{-0.3cm}
ds^2 &=& 
-(1+2A)dt^2+2 \partial_{i} \psi dt dx^i \nonumber \\
\hspace{-0.3cm}
& &+a^2(t) [ (1+2\zeta) \delta_{ij} 
+\gamma_{ij} ]dx^i dx^j\,.
\label{permet}
\ea
The scalar perturbation $E$ appearing as a form 
$\partial_i \partial_j E$ in (\ref{permet}) is set to 0
for fixing the spatial part of gauge-transformation vector. 
We also choose the unitary gauge $\delta \phi=0$ to fix 
the temporal gauge-transformation vector. 

\subsection{Tensor perturbations}

Expanding the action (\ref{act}) up to quadratic order in 
tensor perturbations, the resulting second-order action 
is given by \cite{building,GW15}
\be
S_2^{(h)}=\int d^4 x\,a^3 q_{\rm t} \delta^{ik} \delta^{jl}
\left( \dot{\gamma}_{ij} \dot{\gamma}_{kl}
-\frac{c_{\rm t}^2}{a^2}\partial \gamma_{ij} 
\partial \gamma_{kl} \right)\,, 
\label{L2ten}
\ee
where
\ba
q_{\rm t} &\equiv& 
\frac{L_{,\cal S}}{4}=\frac12 M_{\rm pl}^2F 
\left( 1-\frac{x_4}{5} \right)\,,\\
c_{\rm t}^2 &\equiv& \frac{L_{,\cal R}}{L_{,\cal S}}
=\frac{5+sx_4}{5-x_4}\,,
\label{qtctdef}
\ea
and
\be
s \equiv \frac{b_4}{a_4}\,.
\label{rdef}
\ee
The conditions for avoiding ghosts and Laplacian instabilities 
correspond to $q_{\rm t}>0$ and $c_{\rm t}^2>0$, 
respectively. As we have seen in Sec.~\ref{backsec}, 
the term $x_4$ is much smaller than 1. 
Then, the tensor ghost is absent 
under the condition $F>0$. 
For $s$ at most of the order of 1 (which includes the case 
$s=1/3$ of Horndeski theories), the propagation speed 
squared $c_{\rm t}^2$ is very close to 1 
in the regime $|x_4| \ll 1$. 
The parameter $\alpha_{\rm H}$ defined by 
Eq.~(\ref{alH}) reduces to 
\be
\alpha_{\rm H}=\frac{x_4}{5-x_4} (1-3s)\,,
\label{alHes}
\ee
which is of the order of $x_4$ for $|s| \lesssim O(1)$.
Hence the deviation of $c_{\rm t}^2$ from 1 is the 
same order as $\alpha_{\rm H}$.

\subsection{Propagation speeds of scalar perturbations}

Expansion of the action in GLPV theories up to quadratic order 
in scalar perturbations was carried out in Refs.~\cite{Gergely,KT14,Koyama}. 
In this approach, the perfect fluids of radiation and nonrelativistic matter 
are modeled in terms of k-essence Lagrangians 
$P^{(2)}(\chi_2)=b_r Y_2^2$ \cite{KT14}
and $P^{(3)}(\chi_3)=b_m (Y_3-Y_0)^2$ \cite{Scherrer}, respectively, 
where $b_r$, $b_m$, and $-Y_0$ are positive constants,  
and $Y_2=\nabla_\mu \chi_2 \nabla^\mu \chi_2$ and 
$Y_3=\nabla_\mu \chi_3 \nabla^\mu \chi_3$ are 
the kinetic energies of two scalar fields $\chi_2$ 
and $\chi_3$ \cite{kinetic}. 
Since the corresponding energy densities are given by 
$\rho^{(i)}(\chi_i)=2Y_i P^{(i)}_{,Y_{i}}-P^{(i)}$ (where $i=2,3$), 
it follows that $\rho^{(2)}(\chi_2)=3b_rY_2^2$ and 
$\rho^{(3)}(\chi_3)=b_m(3Y_3+Y_0)(Y_3-Y_0)$. 
Then, the equations of state for radiation and nonrelativistic 
matter are $w_r=P^{(2)}/\rho^{(2)}=1/3$ and 
$w_m=P^{(3)}/\rho^{(3)}=(Y_3-Y_0)/(3Y_3+Y_0)$, respectively.
Provided that $|(Y_3-Y_0)/Y_0| \ll 1$, the matter equation of 
state $w_m$ is close to 0.

The second-order action of Eq.~(\ref{act}) corresponding to 
scalar perturbations is given by $S_2^{(s)}=\int d^4 x\,{\cal L}_2$, 
with the Lagrangian density \cite{Gergely,KT14,KaseIJ}
\be
\mathcal{L}_{2}=a^{3}\left( \dot{\vec{\mathcal{X}}}^{t}{\bm K} 
\dot{\vec{\mathcal{X}}}-\partial _{j}\vec{\mathcal{X}}^{t}{\bm G}
\partial^{j}{\vec{\mathcal{X}}}-\vec{\mathcal{X}}^{t}{\bm B} 
\dot{\vec{\mathcal{X}}}-\vec{\mathcal{X}}^{t}{\bm M} \vec{\mathcal{X}}\right) \,,
\label{L2mat}
\ee
where ${\bm K},{\bm G},{\bm B},{\bm M}$ 
are $3 \times 3$ matrices and the vector $\vec{\mathcal{X}}^{t}$ 
is composed of the curvature perturbation $\zeta$ 
and the perturbations of matter fields, as
$\vec{\mathcal{X}}^{t}=\left( \zeta, \delta \chi_2 /M_{\rm pl},
 \delta \chi_3 /M_{\rm pl} \right)$. 
The nonvanishing components of two matrices ${\bm K}$ and ${\bm G}$, 
which determine the no-ghost and stability conditions in 
the small-scale limit, are given by
\ba
&&
K_{11}=Q_{\rm s}+\frac{K_{12}^2}{K_{22}}+
\frac{K_{13}^2}{K_{33}},~~
K_{ii}=( 2\dot{\chi}_i^2 P_{,Y_iY_i}^{(i)}
-P_{,Y_i}^{(i)} )M_{\rm pl}^2,\nonumber \\
&&
K_{1i}=K_{i1}=-\frac{4L_{,\cal S}\dot{\chi}_i}
{M_{\rm pl}{\cal W}}K_{ii}\,,\nonumber \\
&&
G_{11}=2(\dot{\cal M}+H{\cal M}-L_{,\cal R}),~~~
G_{ii}=c_i^2K_{ii},\nonumber \\
& &
G_{1i}=G_{i1}=-\frac{{\cal M}\dot{\chi}_i}
{L_{,\cal S}M_{\rm pl}}G_{ii}\,,
\ea
where $i=2,3$, and 
\ba
Q_{\rm s} &=& 6L_{,\cal S}+\frac{8L_{,\cal S}^2}{{\cal W}^2}
(2L_{,N}+L_{,NN}-6H{\cal W}+12H^2L_{,\cal S}), \nonumber \\
c_i^2 &=& \frac{P_{,Y_i}^{(i)}}
{P_{,Y_i}^{(i)}-2\dot{\chi}_i^2 P_{,Y_iY_i}^{(i)}}\,,\nonumber \\
{\cal W}&=& L_{,KN}+2HL_{,{\cal S}N}+4L_{,\cal S}H\,, \nonumber \\
{\cal M}&=& \frac{4L_{,\cal S}}{\cal W} 
\left( L_{,N{\cal R}}+L_{,\cal R} \right)\,.
\ea

For the Lagrangian $P^{(2)}(Y_2)=b_r Y_2^2$ of radiation, 
the term $K_{22}=-6b_rY_2$ is positive with the 
propagation speed squared $c_r^2=1/3$. 
For the Lagrangian $P^{(3)}(Y_3)=b_m (Y_3-Y_0)^2$ of 
nonrelativistic matter with $|(Y_3-Y_0)/Y_0| \ll 1$, we obtain 
$K_{33} \simeq -4b_m Y_0 M_{\rm pl}^2>0$ 
with the propagation speed squared 
$c_m^2=(Y_3-Y_0)/(3Y_3-Y_0) \simeq 0$. 
Under the conditions $K_{22}>0$ and $K_{33}>0$, 
the scalar ghost is absent for $Q_{\rm s}>0$, where
\ba
& &
Q_{\rm s}
= \frac{3FM_{\rm pl}^2(5-x_4)}
{25(2-2\sqrt{6}qx_1-x_3-2x_4)^2} \nonumber \\
& & \times \biggl[ 4x_1^2 \{ 5-(1-6q^2)x_4 \}+5x_3(4+x_3)
+12(2+x_3)x_4 \nonumber \\
& &~~~+8x_4^2+4\sqrt{6}qx_1(5x_3+8x_4) \biggr].
\label{Qs}
\ea
Provided that the quantities $|qx_1|,|x_3|,|x_4|$ are much smaller 
than 1, Eq.~(\ref{Qs}) reduces to 
$Q_{\rm s} \simeq 3FM_{\rm pl}^2 (5x_1^2+5x_3+6x_4)/5$. 
Under the no-ghost condition $F>0$ of tensor perturbations,
the condition $Q_{\rm s}>0$ is always satisfied for 
$x_3>0$ and $x_4>0$. 
For the branch $\dot{\phi}>0$, they translate to 
$a_3>0$ and $a_4>0$, respectively.
 
The propagation speeds $c_{s}$ can be derived 
by solving ${\rm det} (c_{s}^2{\bm K}-{\bm G})=0$, i.e., 
\ba
& &\prod_{i=1}^{3} \left( c_{s}^2K_{ii}-G_{ii} \right)
-\left( c_{s}^2K_{12}-G_{12} \right)^2
\left( c_{s}^2K_{33}-G_{33} \right) \nonumber \\
& &-\left( c_{s}^2K_{13}-G_{13} \right)^2
\left( c_{s}^2K_{22}-G_{22} \right)=0\,.
\label{cseq}
\ea
Using the parameter 
$\alpha_{\rm H}=(L_{,N{\cal R}}+L_{,\cal R})/L_{,\cal S}-1
={\cal M}{\cal W}/(4L_{,\cal S}^2)-1$, we obtain the 
relations $G_{12}=K_{12} (1+\alpha_{\rm H})c_r^2$ 
and $G_{13}=K_{13} (1+\alpha_{\rm H})c_m^2$ 
with $c_r^2=G_{22}/K_{22}$ and $c_m^2=G_{33}/K_{33}$. 
We also define the following quantities 
\ba
& &c_{\rm H}^2=\frac{1}{Q_{\rm s}} \left( G_{11} 
-\frac{K_{12}^2}{K_{22}}c_r^2-\frac{K_{13}^2}{K_{33}}c_m^2 \right)\,,\label{cH}\\
& &\beta_r=\frac{K_{12}^2}{K_{22}} \frac{2c_r^2\alpha_{\rm H}}{Q_{\rm s}}\,,
\quad
\beta_m=\frac{K_{13}^2}{K_{33}} \frac{2c_m^2\alpha_{\rm H}}{Q_{\rm s}}\,,\\
& &\beta_{\rm H}=\beta_r+\beta_m\,.
\label{betaH}
\ea

In Horndeski theories ($\alpha_{\rm H}=0$), the solutions to Eq.~(\ref{cseq}) are given by 
$c_s^2=c_{\rm H}^2$, $c_s^2=c_r^2$, and $c_s^2=c_m^2$. 
In GLPV theories the propagation speeds are generally mixed each other.
For the case $c_m^2=0$, the matter sound speed $\tilde{c}_m$ is decoupled from other propagation speeds, such that one of the solutions is given by $c_s^2=0$. 
Other solutions to Eq.~(\ref{cseq}) are expressed as
\ba
c_{\rm s}^2 &=&
\frac12 \left[ c_r^2+c_{\rm H}^2-\beta_{\rm H}-
\sqrt{(c_r^2-c_{\rm H}^2+\beta_{\rm H})^2+
2c_r^2\alpha_{\rm H}\beta_r} \right],\nonumber \\
\tilde{c}_{r}^2 &=&
\frac12 \left[ c_r^2+c_{\rm H}^2-\beta_{\rm H}+
\sqrt{(c_r^2-c_{\rm H}^2+\beta_{\rm H})^2+
2c_r^2\alpha_{\rm H}\beta_r} \right].
\nonumber\\
\ea
In the regime $|\alpha_{\rm H}| \ll 1$ these reduce to 
$c_{\rm s}^2 \simeq c_{\rm H}^2-\beta_{\rm H}
+\alpha_{\rm H} \beta_{r}c_r^2/
[2(c_{\rm H}^2-c_r^2-\beta_{\rm H})]$ and 
$\tilde{c}_{r}^2 \simeq c_{r}^2-\alpha_{\rm H}
\beta_{r}c_r^2/
[2(c_{\rm H}^2-c_r^2-\beta_{\rm H})]$, respectively. 
Since the term $\beta_{\rm H}$ is not necessarily small 
even for $|\alpha_{\rm H}| \ll 1$ the deviation from 
Horndeski theories generally gives rise to a non-negligible 
contribution to $c_{\rm s}^2$, while the modification to 
$\tilde{c}_r^2$ can be negligible \cite{Koyama}. 

Let us estimate $c_{\rm s}^2$ during the radiation-dominated 
epoch ($x_5^2 \simeq 1$) under the condition that 
$x_4 \gg \{ x_1^2,x_3 \}$ with $x_i \ll 1$ (where $i=1,2,3,4$). 
Since $\ddot{\phi}/(H\dot{\phi}) \simeq 
1/3-5\sqrt{6}qx_1(1-x_5^2)/(12x_4)$ from 
Eq.~(\ref{ddotphi2}), we can estimate the quantities 
defined in Eqs.~(\ref{cH}) and (\ref{betaH}), as
$c_{\rm H}^2 \simeq (4-7s)/9-5\sqrt{6}qx_1(1-s)(1-x_5^2)
/(18x_4)$ and $\beta_{\rm H} \simeq 2(1-3s)/9$, respectively. 
Hence the scalar propagation speed squared 
is approximately given by 
\be
c_{\rm s}^2 \simeq \frac19 (2-s)
-\frac{5\sqrt{6}qx_1(1-s)(1-x_5^2)}{18x_4}\,.
\label{csrad}
\ee
To avoid the instability of scalar perturbations, the positivity of 
the first term on the rhs of Eq.~(\ref{csrad}) requires that 
$s \leq 2$. As $x_5^2$ starts to decrease from 1, 
the second term on the rhs of Eq.~(\ref{csrad}) is not necessarily 
negligible for $x_4 \ll x_1$ (which can happen even 
for $x_4 \gg x_1^2$).
This second term does not become negative for $s \leq 1$, so 
the sufficient condition for avoiding the Laplacian 
instability is given by  
\be
s \leq 1\,.
\label{rcon}
\ee
\begin{figure}
\includegraphics[height=3.2in,width=3.3in]{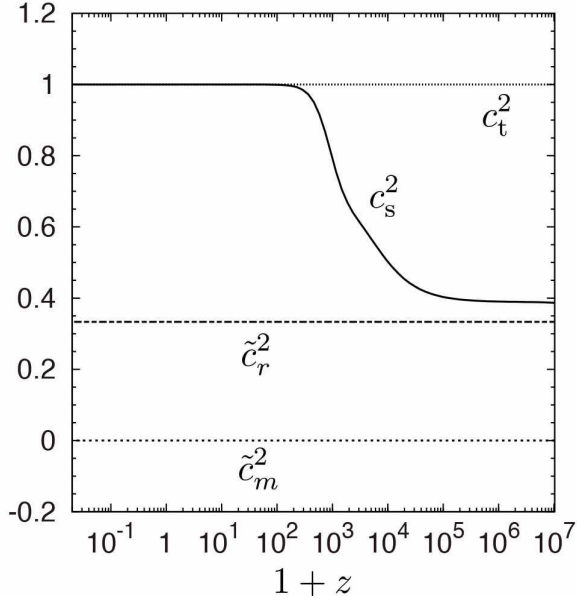}
\caption{\label{csfig}
Evolution of $c_{\rm t}^2$, $c_{\rm s}^2$, 
$\tilde{c}_r^2$, and $\tilde{c}_m^2$ versus $1+z$ 
for the same model parameter and initial conditions 
as those used in Fig.~\ref{xifig}. 
The ratio $s=b_4/a_4$ is chosen to be $s=1/2$.}
\end{figure}
\begin{figure}
\includegraphics[height=3.2in,width=3.3in]{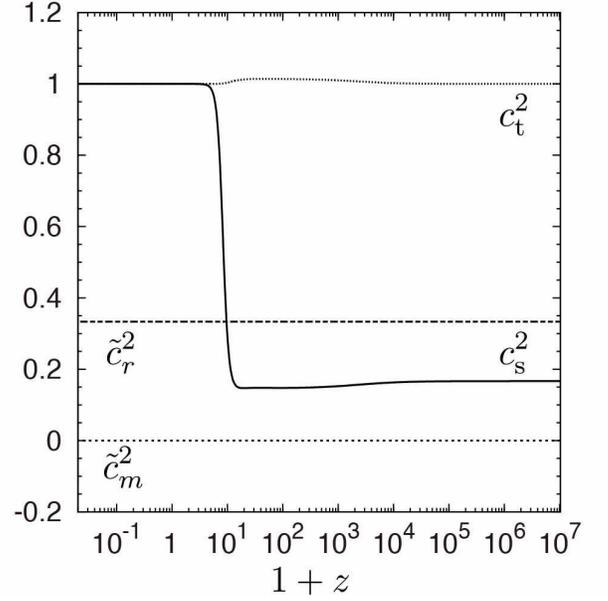}
\caption{\label{csfig2}
Evolution of $c_{\rm t}^2$, $c_{\rm s}^2$, 
$\tilde{c}_r^2$, and $\tilde{c}_m^2$ versus $1+z$ 
for the same model parameter and initial conditions 
as those used in Fig.~\ref{xifig2}, with $s=1/2$.}
\end{figure}

In Fig.~\ref{csfig} we plot the evolution of $c_{\rm s}^2$ as well as $c_{\rm t}^2$, $\tilde{c}_r^2$, and $\tilde{c}_m^2$ for $s=1/2$ and the model parameters same as those used 
in Fig.~\ref{xifig}. 
The scalar propagation speed squared starts to evolve from 
the value $c_{\rm s}^2 \simeq 0.387$ in the deep radiation era, 
which shows good agreement with the analytic estimation (\ref{csrad}). 
In this case the quantity $x_4$ is initially much larger than 
$x_1^2$, but after $x_4$ is caught up with $x_1^2$, 
it becomes negligibly small relative to $x_1^2$ 
(see Fig.~\ref{xifig}).
Once $x_1^2$ dominates over $x_4$ and $x_3$, 
$c_{\rm s}^2$ approaches 1. 
As we see in Fig.~\ref{csfig}, $\tilde{c}_r^2$ is very close to 
$c_r^2=1/3$, which means that the correction to $\tilde{c}_r^2$ 
arising from the deviation from Horndeski theories is small. 
As we estimated in Eqs.~(\ref{qtctdef}) and (\ref{alHes}), 
the departure of $c_{\rm t}^2$ from 1 is of 
the order of $\alpha_{\rm H}$. 

For larger initial values of $x_4$, it happens that 
$x_4$ dominates over $x_1^2$ 
even in the deep matter era (see Fig.~\ref{xifig2}). 
In this epoch the analytic estimation shows that 
$c_{\rm H}^2 \simeq [11-17s+(5-11s)x_5^2]/36
+5\sqrt{6}q(s-1)x_1/(18x_4)$ and 
$\beta_{\rm H} \simeq (3\Omega_m+4x_5^2)(1-3s)/18$, 
so $c_{\rm s}^2$ is approximately given by 
\ba
c_{\rm s}^2 &\simeq& \frac{1}{36} \left[11-17s+6\Omega_m
(3s-1)+(13s-3)x_5^2 \right] \nonumber \\
&& +\frac{5\sqrt{6}q(s-1)x_1}{18x_4}\,.
\label{csmat}
\ea
In Fig.~\ref{csfig2} we illustrate the evolution of the propagation 
speeds for the initial conditions used in Fig.~\ref{xifig2}, with $s=1/2$.
Since the condition $x_4 \gg \{ x_1^2,x_3 \}$ is satisfied by the redshift 
$z \approx 10$, $c_{\rm s}^2$ starts to evolve from the value (\ref{csrad}) 
in the radiation era to the value (\ref{csmat}) in the deep matter era. 
Indeed, our numerical results are in good agreement with 
the analytic estimations (\ref{csrad}) and (\ref{csmat}). 
After $x_4$ becomes sub-dominant to $x_1^2$, 
$c_{\rm s}^2$ approaches 1. 
Since $x_4$ is quite large during the deep matter era, 
this also leads to some deviation of $c_{\rm t}^2$ from 1. 

{}From Eq.~(\ref{csmat}) we find that $c_{\rm s}^2$ becomes negative 
for $s$ largely negative. Our numerical simulations show that, 
for $s<0$, $c_{\rm s}^2$ tends to decrease around the epoch 
$x_4 \approx x_1^2$ and it temporally enters the region with 
negative $c_{\rm s}^2$. 
After $c_{\rm s}^2$ reaches a minimum, 
it begins to increase again to approach the value $1$.
Combining the condition $s \geq 0$ with Eq.~(\ref{rcon}), 
the Laplacian instability of scalar perturbations can be avoided for
\be
0 \leq s \leq 1\,,
\label{scon}
\ee
which includes Horndeski theories as a special case.

If $x_3$ dominates over $x_1^2$ and $x_4$ during the radiation 
and the deep matter eras, we obtain the approximate relations 
$c_{\rm H}^2 \simeq (5+x_5^2)/12-\sqrt{6}qx_1(1-x_5^2)/(6x_3)$ 
and $\beta_{\rm H} \simeq (4x_5^2+3\Omega_m)(1-3s)x_4/(15x_3)$. 
Since $|\beta_{\rm H}| \ll 1$, we have that  
$c_{\rm s}^2 \simeq c_{\rm H}^2$.
Under the no-ghost condition $x_3>0$, 
it follows that $c_{\rm s}^2>0$.
Hence this case does not provide additional 
constraints on $s$.

\subsection{Evolution of cosmological perturbations}
\label{matterpersec}

We study the evolution of cosmological perturbations after the 
onset of the matter-dominated epoch to confront the model 
with observations of large-scale structures and weak lensing.
In doing so, we introduce the perturbations of energy density 
and momentum of pressureless matter, as 
$\delta T^0_0=-\delta \rho$ and $\delta T^0_i=\partial_i \delta q$. 
The continuity equations ${\delta T^{\mu}}_{0;\mu}=0$ and 
${\delta T^{\mu}}_{i;\mu}=0$ give rise to the linear perturbation 
equations of motion in Fourier space, as
\ba
&&
\dot{\delta \rho}+3H\delta \rho=-\rho 
\left( 3\dot{\zeta}+\frac{k^2}{a^2}\psi \right)
+\frac{k^2}{a^2} \delta q\,,
\label{mattereq1}\\
&&
\dot{\delta q}+3H \delta q=-\rho A\,,
\label{mattereq2}
\ea
respectively, where $k$ is a comoving wave number. 
The gauge-invariant density contrast is defined by 
\be
\delta_m \equiv \delta-3V_m\,,
\ee
where $\delta \equiv \delta \rho/\rho_m$ and 
$V_m \equiv H \delta q/\rho_m$. 
Differentiating Eq.~(\ref{mattereq1}) with respect to $t$ 
and using Eq.~(\ref{mattereq2}), we obtain 
\be
\ddot{\delta}_m+2H\dot{\delta}_m+\frac{k^2}{a^2}\Psi
=-3\ddot{\cal B}-6H\dot{\cal B}\,,
\ee
where ${\cal B} \equiv \zeta+V_m$, and $\Psi$ corresponds 
to the gauge-invariant gravitational potential 
given by \cite{Bardeen}
\be
\Psi \equiv A+\dot{\psi}\,.
\ee
We introduce another gauge-invariant potential 
\be
\Phi \equiv \zeta+H\psi\,.
\ee
The gravitational slip parameter characterizing the 
difference between $-\Psi$ and $\Phi$ is defined by 
$\gamma \equiv -\Phi/\Psi$.

The effective gravitational coupling $G_{\rm eff}$ 
between $\Psi$ and $\delta_m$ is introduced as
\be
\frac{k^2}{a^2}\Psi=-4\pi G_{\rm eff} \rho_m \delta_m\,.
\label{Psieq}
\ee
In order to know $G_{\rm eff}$ and the growth rate 
of $\delta_m$ explicitly, we need to solve other perturbation equations of motion. 
The full linear perturbations of motion in GLPV theories are 
given in Refs.~\cite{building,Gergely,Tsuji15}. Since we are primarily interested in the 
evolution of perturbations in the low-redshift regime, 
we neglect the contribution of radiation in the following discussion.

In our model, the perturbations $\zeta$, $\chi \equiv H \psi$, 
$\delta$, and $V_m$ obey
\begin{widetext}
\ba
\hspace{-0.9cm}
&&\zeta' = \frac{5(3\Omega_m V_m+ \mu A)}{2(5-x_4)}\,,
\label{zetaeq} \\
\hspace{-0.9cm}
&&\chi'= \frac{3[5(h-1+2\sqrt{6}qx_1)
+x_4(1-h+4\epsilon_{\phi})]\chi-3(5-x_4)(1+\alpha_{\rm H})A
-[15+x_4-(5-x_4)\alpha_{\rm H}]\zeta}{3(5-x_4)}\,,\\
\hspace{-0.9cm}
&&\delta'=-\frac{15(3\Omega_m V_m+\mu A)}{2(5-x_4)}
+{\cal K}^2 (V_m-\chi),\\
\hspace{-0.9cm}
%&&\delta'= \frac{-15\Omega_m \delta-30 
%[1-x_1\{x_1+2q(\sqrt{6}-3qx_1)\}-2x_3-3x_4]A
%+2{\cal K}^2(1+\alpha_{\rm H})(5-x_4)\zeta}{5\mu}
%+{\cal K}^2 V_m, \\
%\hspace{-0.9cm}
&&V_m' = hV_m-A\,,\label{Vmeqd}\\
\hspace{-0.9cm}
&&A= \frac{2(5-x_4)[15\Omega_m \delta-{\cal K}^2
\{ 5\mu \chi+2(5-x_4)(1+\alpha_{\rm H})\zeta \}]-225\mu\Omega_m V_m}
{60x_1^2[5-(1-6q^2)x_4]+75x_3(4+x_3)+180(2+x_3)x_4
+120x_4^2+60\sqrt{6}qx_1(5x_3+8x_4)},
\label{Aeq}
\ea
\end{widetext}
where $\mu \equiv 2-2\sqrt{6} qx_1-x_3-2x_4$, $k$ is 
a coming wave number, and ${\cal K} \equiv k/(aH)$. 
Note that we have also used the relation 
$sx_4=[(1+\alpha_{\rm H})x_4-5\alpha_{\rm H}]/3$.

\begin{figure}
\includegraphics[height=3.2in,width=3.3in]{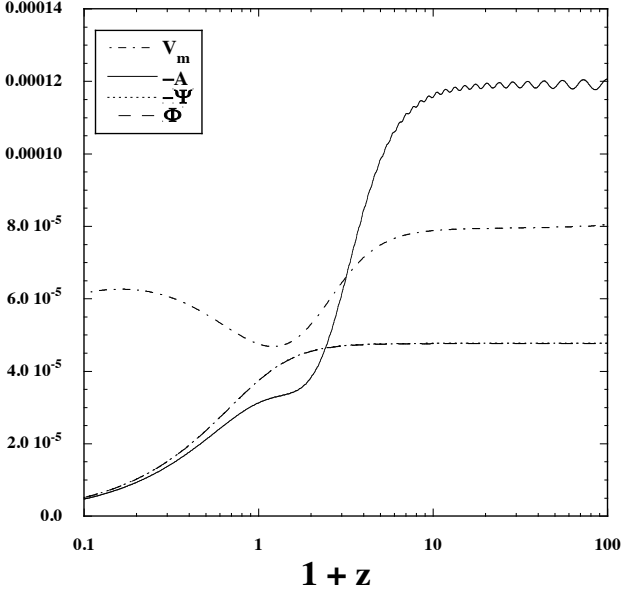}
\caption{\label{grafig}
Evolution of the perturbations $V_m$, $-A$, $-\Psi$, and $\Phi$ for 
$q=-0.02$ and $s=1/2$ with the initial conditions 
$x_1=1.60 \times 10^{-2}$,  $x_2=2.30 \times 10^{-6}$, $x_3=0$, 
$x_4=1.00 \times 10^{-11}$, 
$\lambda=0.31$, $\zeta=-1.02 \times 10^{-6}$, 
$\chi=4.86 \times 10^{-5}$, $\delta=1.05 \times 10^{-2}$, 
$V_m=8.04 \times 10^{-5}$
and ${\cal K}=18$ at $z=99.5$. 
In this case, the normalized wave number  
corresponds to ${\cal K}_0=103$ today.}
\end{figure}

Let us derive the second-order equation for $V_m$ under 
the condition that $x_1^2 \gg \{ x_3,x_4 \}$ while
keeping the term $\alpha_{\rm H}$. 
Taking the ${\cal N}$ derivative of Eq.~(\ref{Vmeqd}) 
and using other equations of motion, we obtain 
\ba
V_m''+\alpha_1 V_m'+\alpha_2 V_m=
\beta_1 \chi+\beta_2\zeta\,,
\label{Vmeq}
\ea
where $\alpha_1$ and $\alpha_2$ are time-dependent 
coefficients, and 
\ba
\hspace{-0.8cm}
& &
\beta_1=-\frac{(3x_1+\sqrt{6}q-6q^2x_1)
{\cal K}^2}{3x_1}\,,\\
\hspace{-0.8cm}
& &
\beta_2=-\frac{\sqrt{6} [ 6qx_1 (3+7\alpha_{\rm H})
-\sqrt{6} (4\alpha_{\rm H}+3\alpha_{\rm H}')]
{\cal K}^2}{54x_1^2}.
\ea

The term $\alpha_2$ contains the $k$-dependent contribution:
\be
\alpha_2 \supset -\alpha_{\rm H} 
\frac{\Omega_m}{2x_1^2}{\cal K}^2 
\equiv m_{\alpha_{\rm H}}^2\,.
\ee
The parameter $\alpha_{\rm H}$ is of the order of $x_4$, 
so the coefficient $-\alpha_{\rm H}\Omega_m/(2x_1^2)$ in 
$m_{\alpha_{\rm H}}^2$ is smaller than the order of 1.
This property is different from that studied in the model 
of Ref.~\cite{Koyama}, where 
$\alpha_{\rm H}$ is not necessarily small. 
In the latter model, the mass squared $m_{\alpha_{\rm H}}^2$ 
can induce an oscillation of $V_m$ with a large frequency. 
To avoid this heavy oscillation,
the initial conditions of $V_m$ need to be fine-tuned as 
$V_m=(\beta_1 \chi+\beta_2 \zeta)/\alpha_2$ and $V_m'=0$.
Otherwise, a rapid oscillation of the gravitational potential 
$\Psi$ with its sign change is induced \cite{Koyama}. 
This rapid oscillation of $\Psi$ can contradict the CMB observations.

In our model, the term $m_{\alpha_{\rm H}}^2$ is not much larger than 
1 for sub-horizon perturbations associated with large-scale structures, 
so we can avoid the appearance of  heavy oscillations of $\Psi$.
In Fig.~\ref{grafig} we plot the evolution of the perturbations 
$V_m$ and $-A$ as well as the gravitational potentials 
$-\Psi$ and $\Phi$ for the background initial conditions 
similar to those used in Fig.~\ref{xifig} but with $x_3=0$.
The initial conditions of perturbations are chosen to satisfy 
$\zeta' \simeq 0$, $\chi' \simeq 0$, and 
$V_m \simeq (\beta_1 \chi+\beta_2 \zeta)/\alpha_2$. 
Since the derivative $V_m'$ initially differs from 0, 
this choice does not correspond to the fine-tuned 
initial conditions employed in Ref.~\cite{Koyama}. 

In Fig.~\ref{grafig} we find that the gauge-invariant gravitational potentials 
$-\Psi$ and $\Phi$ stay nearly constant without oscillations. 
The perturbation $A=hV_m-V_m'$ exhibits 
a tiny oscillation due to the derivative term $V_m'$, 
but this hardly affects the regular behavior of 
$\Psi$. Choosing non-zero initial values of $x_3$ as 
those used in Fig.~\ref{xifig}, the effect of the 
term $x_3$ can even eliminate the small oscillation of $A$.
Recall that the sound speed squared 
$c_{\rm s}^2$ associated with the curvature 
perturbation $\zeta$ is at most of the order of 1, 
so the gravitational potential $\Phi$ does not 
possess a large effective mass.
We also have the following exact relation 
\ba
\hspace{-0.7cm}
& &
\Psi+\Phi \nonumber \\
\hspace{-0.7cm}
&&=\frac{2\chi(5\sqrt{6}qx_1+2x_4 \epsilon_{\phi})
-\zeta x_4(s+1)+Ax_4(3s-1)}{5-x_4}. 
\label{PsiPhi} \nonumber \\
\ea
As long as $x_4 \ll 1$ and $q x_1 \ll 1$, the gravitational 
slip parameter $\gamma$ is close to 1. 
As we see in Fig.~\ref{grafig}, the evolution of $-\Psi$ is 
very similar to that of $\Phi$.

Figure \ref{grafig} corresponds to the case in which 
$x_1^2$ dominates over $x_4$ after the end of 
the radiation era, but it also happens that 
$x_4 \gg x_1^2$ in the deep {\it matter} 
era for larger initial values of $x_4$ 
(such as the case of Fig.~\ref{xifig2}). 
In the latter case, the term $\alpha_2$ in Eq.~(\ref{Vmeq}) 
involves the contribution 
\be
\alpha_2 \supset -\alpha_{\rm H} 
\frac{5\Omega_m}{12x_4}{\cal K}^2\,,
\ee
whose coefficient in front of ${\cal K}^2$ is smaller 
than the order of 1. Hence the heavy oscillating mode 
does not arise in this case as well. 
This is consistent with the estimation of $c_{\rm s}^2$ 
given in Eq.~(\ref{csmat}).

\begin{figure}
\includegraphics[height=3.2in,width=3.4in]{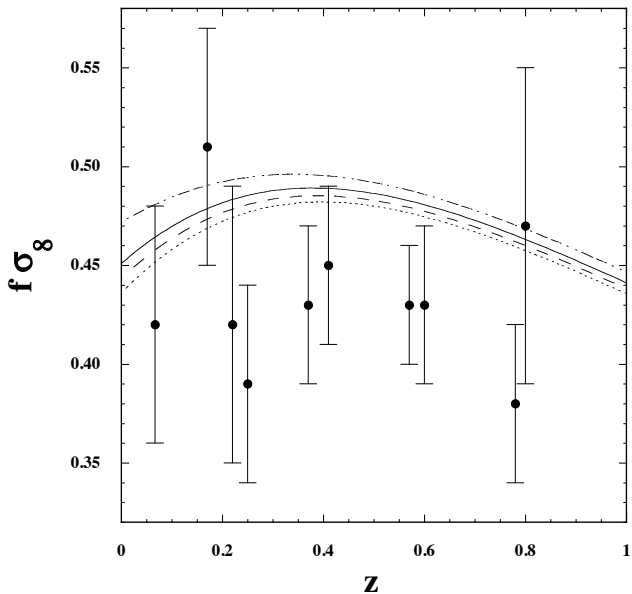}
\caption{\label{fsigmafig}
Theoretical predictions of $f(z)\sigma_8(z)$ 
for $q=-0.15, -0.1, -0.07, -0.01$ 
(from top to bottom) and $s=1/2$. 
The perturbations correspond to the today's wave
number ${\cal K}_0=100$ with $\sigma_8(0)=0.82$.
The initial conditions are chosen to be $x_1=1.60 \times 10^{-2}$, 
$x_2=2.20 \times 10^{-6}$, $x_3=1.60 \times 10^{-7}$,  
$x_4=1.10 \times 10^{-11}$, and $\lambda=0.31$ around the 
redshift $z=100$. The black points with error bars represent the 
observational data of $f(z)\sigma_8(z)$ constrained from 
6dFGRS \cite{6dF}, 2dFGRS \cite{2dF}, WiggleZ \cite{Wiggle}, SDSSLRG \cite{SDSS}, BOSS- CMASS \cite{BOSS}, 
and VIPERS \cite{VIPERS} surveys.}
\end{figure}

The growth rate $\delta_m'$ of matter perturbations can be 
measured from the observations of red shift-space distortions 
(RSD) at low redshifts ($z<O(1)$) \cite{Kaiser}. 
In this regime, the gravitational potentials $-\Psi$ and $\Phi$ 
start to decrease due to the onset of cosmic acceleration 
(see Fig.~\ref{grafig}). Numerically, we solve 
Eqs.~(\ref{zetaeq})-(\ref{Aeq}) to find the evolution 
of the matter density contrast $\delta_m=\delta-3V_m$.
The effective gravitational coupling $G_{\rm eff}$ is
known from Eq.~(\ref{Psieq}), such that 
\be
{\cal K}^2 \Psi=-\frac32 \frac{G_{\rm eff}}{G}
F\Omega_m \delta_m\,,
\ee
where $G=(8\pi M_{\rm pl}^2)^{-1}$.

Provided that the contributions from the terms $x_3$ and 
$x_4$ to the perturbation equations of motion are 
negligible to those involving the coupling $q$, 
the quasi-static approximation 
on sub-horizon scales gives rise to the 
gravitational coupling \cite{Yoko}
\be
G_{\rm eff} \simeq \frac{G}{F} \left( 1+2q^2 \right)\,.
\label{Geffes}
\ee
Numerically we have confirmed that $G_{\rm eff}$ is well 
described by Eq.~(\ref{Geffes}) at low redshifts. 

In Fig.~\ref{fsigmafig} we show the evolution of 
$f(z)\sigma_8(z)$ for four different values 
of $q$, where $f \equiv \dot{\delta}_m/(H \delta_m)$ and 
$\sigma_8$ is the amplitude of over-density at the comoving 
$8\,h^{-1}$ Mpc scale ($h$ is the normalized Hubble constant). 
For increasing $|q|$, $f(z)\sigma_8(z)$ gets 
larger, but for the coupling $|q| \lesssim 0.1$, 
the theoretical curves of $f(z)\sigma_8(z)$ are not significantly 
different from each other.
This reflects the fact that the term $1+2q^2$ in Eq.~(\ref{Geffes}) 
is close to 1 for $|q|$ smaller than the order of 0.1. 
As long as $\{x_3,x_4\} \ll qx_1$ in the regime $z<O(1)$, 
the theoretical prediction of $f(z)\sigma_8(z)$ 
for given $q$ is insensitive to the choice of initial conditions.

In Fig.~\ref{fsigmafig} we have chosen the Planck best-fit 
value $\sigma_8(0)=0.82$, in which case there is a tension 
between the theoretical curves and the RSD data.
This tension, which is also present for the concordance 
cosmological model \cite{Planckmo}, comes from the fact 
that the CMB observation 
favors $\sigma_8(0)$ larger than those
constrained from low-redshift measurements. 
The current RSD data alone are not sufficient 
to put trustable bounds on $q$.

%===========================================
%section IV
\section{Vainshtein screening}
\label{vasec} 
%===========================================

We study how the fifth force mediated by the field $\phi$ can be suppressed on the spherically symmetric and static background described by the metric
\be
ds^{2}=-e^{2\Psi(r)}dt^{2}+e^{2\Phi(r)}dr^{2}
+r^{2} (d\theta^{2}+\sin^{2}\theta\, 
d\varphi^{2})\,.
\label{line}
\ee
The gravitational potentials $\Psi(r)$ and $\Phi(r)$ are
functions of the distance $r$ from the center of symmetry 
with a matter source.
Considering a situation in which the dominant contributions 
to the background equations of motion are of the order of 
$A_4\Psi/r^2$ and $A_4\Phi/r^2$, we can derive the equation 
of motion for the scalar field $\phi$ under the approximation of 
weak gravity ($|\Psi| \ll 1, |\Phi| \ll 1$). 
In GLPV theories this is given by equation 
(6.1) of Ref.~\cite{DKT15} (see also 
Refs.~\cite{DKT11,Kase13} for the same approximation and 
Refs.~\cite{Koba15,Sakstein,Mizuno,Jimenez} 
for related works).

In the presence of the term $a_3X$ in $C_3$ and the terms 
$a_4X^2, b_4X^2$ in $A_4, B_4$, the latter contributions 
to the field profile
dominate over the former for the distance $r$ associated with 
the recovery of GR \cite{Kase13}. Hence, we set $a_3=0$ 
in the following and discuss how the Vainshtein mechanism works with 
the terms $a_4X^2$ and $b_4X^2$.
Then, the field equation of motion approximately reads
\be
\frac{1}{r^2} \frac{d}{dr} \left[ r^2 \phi'(r) \right]
\simeq \mu_1 \rho_m+\mu_2\,,
\label{phieq}
\ee
where a prime in this section represents a derivative 
with respect to $r$, and 
\ba
\hspace{-0.5cm}
\mu_1 &=&
-\frac{(qM_{\rm pl}F-\beta \phi')r}{2\beta 
(M_{\rm pl}^2 F-2a_4{\phi'}^4)}\,,
\label{mu1def} \\
\hspace{-0.5cm}
\mu_2 &=&
\frac{1}{\beta r} \left[ \left( \frac14 \omega_{,\phi}{\phi'}^2
-\frac12 a_1 \right)r^2-\frac{24}{r}(a_4-b_4){\phi'}^3 \right],\\
\hspace{-0.5cm}
\beta &=& -\frac12 \omega r-\frac{12}{r}(a_4-b_4){\phi'}^2\,. 
\label{betadef}
\ea

The Vainshtein radius $r_V$ is defined by the distance at which 
the two terms on the rhs of Eq.~(\ref{betadef}) become comparable to each other, i.e., 
\be
r_V^2=24\left|\frac{a_4(1-s)}{\omega(r_V)}\right| {\phi'}^2(r_V)\,.
\label{rV}
\ee
In the regime $r>r_V$, neglecting the $\phi'$-dependent terms in Eqs.~(\ref{mu1def})-(\ref{betadef}),  
it follows that $\mu_1 \simeq q/(\omega M_{\rm pl})$
and $\mu_2 \simeq a_1/\omega$ with 
$\beta \simeq -\omega r/2$.
Under the condition that $\omega$ stays nearly constant, integration of Eq.~(\ref{phieq}) leads to the following solution 
\be
\phi'(r)=\frac{qM_{\rm pl}r_g}{\omega_0 r^2}
+\frac{a_1 r}{3\omega_0}\,,
\label{phiso}
\ee
where $\omega_0 \equiv \omega(r_V)$, and 
$r_g \equiv M_{\rm pl}^{-2} \int \rho_m (\tilde{r})\,\tilde{r}^2 d\tilde{r}$ 
is the Schwarzschild radius of the source.
The second term on the rhs of Eq.~(\ref{phiso}) dominates 
over the first one for the distance $r$ larger than $r_*$, where 
\be
r_*= \left( \frac{3|q|M_{\rm pl}r_g}{|a_1|} \right)^{1/3}\,.
\label{rstar}
\ee
In the regime $r>r_*$ the approximation of ignoring the 
$\phi'$-dependent terms in 
Eqs.~(\ref{mu1def})-(\ref{betadef}) 
is no longer justified, so the solution (\ref{phiso}) is valid 
only for $r<r_*$. In fact, for the radius $r>r_*$, 
the expansion of the Universe comes into play, so the assumption of the static background (\ref{line}) 
starts to lose its validity.

For $r_V<r \ll r_*$, the solution (\ref{phiso}) is approximately given by $\phi'(r) \simeq qM_{\rm pl}r_g/(\omega_0 r^2)$. 
On using this solution, 
Eq.~(\ref{rV}) gives the Vainshtein radius
\be
r_V=\frac{(|q|M_{\rm pl}r_g)^{1/3}}
{\omega_0^{1/2}M}\,,
\label{rVes}
\ee
where 
\be
M \equiv [24a_4(1-s)]^{-1/6}\,.
\ee
To avoid that $r_V$ goes to 0, we require the condition
$s \neq 1$. Combining this with Eq.~(\ref{scon}), 
the parameter $s$ needs to be in the range
\be
0 \le s <1\,.
\ee
We also obtain the field profile 
$\phi(r)=\phi_0-qM_{\rm pl}r_g/(\omega_0r)$, where 
$\phi_0$ is an integration constant. Provided $|\phi_0|$ 
is of the order of $M_{\rm pl}$, it follows that 
$\phi \simeq \phi_0$ for $r \gg r_g$.
Hence the assumption that the term $\omega(\phi)=(1-6q^2)F(\phi)$ 
stays nearly constant is justified. In the following, we employ 
the approximation that the terms $F$ and $\omega$ are 
constants of the order of 1.

Since the potential $V(\phi)=a_1\phi$ drives the present cosmic 
acceleration, we have $M_{\rm pl}^2H_0^2 \simeq a_1\phi 
\simeq |a_1|M_{\rm pl}$ and hence $|a_1| \simeq M_{\rm pl} H_0^2$, 
where $H_0$ is the today's Hubble parameter. 
Then, the distance (\ref{rstar}) can be estimated as 
$r_* \simeq (|q|r_gH_0^{-2})^{1/3}$. 
Defining the ratio 
\be
\xi_M \equiv \frac{M^3}{H_0^2M_{\rm pl}}\,,
\ee
the Vainshtein radius (\ref{rVes}) reads
\be
r_V \simeq \left( \frac{|q|}{\xi_M} 
r_g H_0^{-2} \right)^{1/3}\,.
\label{rVes2}
\ee
As long as $\xi_M \gg 1$, we have that $r_V \ll r_*$ 
and hence the discussion given above is consistent.

For the distance $r<r_V$ the term $\mu_2 \simeq 2\phi'/r$ 
dominates over $\mu_1 \rho_m$ 
with $\beta \simeq -12(a_4-b_4){\phi'}^2/r$, so the solution to 
Eq.~(\ref{phieq}) is simply given by $\phi'(r)={\rm constant}$.
Matching this solution with the one in the regime $r_V<r \ll r_*$, 
we obtain 
\be
\phi'(r)=\frac{qM_{\rm pl}r_g}{\omega_0 r_V^2}\,,
\label{phiso2}
\ee
which is valid for $r_g \ll r<r_V$. 
The differential equations for the gravitational potentials 
$\Phi$ and $\Psi$ in the aforementioned scheme of 
approximation are given by equations (6.7) and (6.8) 
of Ref.~\cite{DKT15}. On using the approximation
$A_4^{-1} \simeq -2[1+2a_4{\phi'}^4/(M_{\rm pl}^2F_0)]
/(M_{\rm pl}^2F_0)$ and the solution (\ref{phiso2}), 
the integrated solutions to $\Phi$ and $\Psi$ read
\ba
\hspace{-0.7cm}
& &\Phi (r)\simeq 
\frac{r_g}{2F_0r} \biggl[ 1-\frac{2q^2}{1-6q^2} 
\left( \frac{r}{r_V} \right)^2+\frac{a_1\phi\,r^3}
{3M_{\rm pl}^2r_g} \nonumber \\
\hspace{-0.7cm}
&&+\frac{q^2(1+2q^2)r_gr^3}{6F_0(1-6q^2)^2r_V^4}
-\frac{2a_4(1+s)M_{\rm pl}^2q^4r_g^3r}
{F_0^4(1-6q^2)^4r_V^8} \biggr]\,,\label{Phiso}\\
\hspace{-0.7cm}
&&\Psi (r) \simeq
-\frac{r_g}{2F_0r} \biggl[ 1-\frac{2q^2}{1-6q^2} 
\left( \frac{r}{r_V} \right)^2+\frac{a_1\phi\,r^3}
{3M_{\rm pl}^2r_g} \nonumber \\
\hspace{-0.7cm}
&&-\frac{q^2(1-4q^2)r_gr^3}{3F_0(1-6q^2)^2r_V^4}
-\frac{8a_4(1-s)M_{\rm pl}^2q^4r_g^3r}
{F_0^4(1-6q^2)^4r_V^8}\ln \frac{r}{r_{c}} 
\biggr]\,,
\label{Psiso}
\ea
where $r_{c}$ is an integration constant.

The leading-order contributions to Eqs.~(\ref{Phiso}) and (\ref{Psiso}) are
$\Phi_{\rm lead}(r)=r_g/(2F_0r)$ and $\Psi_{\rm lead}(r)=-r_g/(2F_0r)$, 
respectively. Other terms are corrections to the leading-order terms, so 
the gravitational slip parameter $\gamma=-\Phi/\Psi$ reads
\ba
\hspace{-0.8cm}
\gamma 
&\simeq& 
1+\frac{q^2 (1-2q^2)r_gr^3}{2F_0(1-6q^2)^2r_V^4}
\nonumber \\
\hspace{-0.8cm}
& &-\frac{2M_{\rm pl}^2q^4r_g^3r}
{F_0^2 (1-6q^2)^4r_V^8}
a_4 \left[1+s-4(1-s)\ln \frac{r}{r_{c}} \right].
\label{gammaes}
\ea
The maximum deviation of $\gamma$ from 1 occurs around 
$r=r_V$. Substituting $r=r_V$ into the second term 
$\Delta \gamma_1$ on the rhs of Eq.~(\ref{gammaes}), 
we find that $\Delta \gamma_1 \simeq q^2r_g/r_V$. 
If $r_V$ is of the order of the 
solar-system scale $r_{\oplus} \simeq 10^{15}$\,cm,  
we have that $\Delta \gamma_1 \simeq 10^{-10} q^2$, 
where we used $r_g \simeq 3 \times 10^{5}$\,cm of the Sun.
Similarly, using Eq.~(\ref{rVes}) with $M^6 \simeq O(0.1)\,a_4^{-1}$ 
and the fact that $\ln r/r_{c}$ is at most of the order 10 for 
$r_{c}$ between $r_g$ and $r_V$, it follows that the third term 
$\Delta \gamma_2$ on the rhs of Eq.~(\ref{gammaes}) 
is also of the order of $q^2r_g/r_V$. 
Hence, for $|q| \lesssim 1$, the local gravity bound
$|\gamma-1|<2.3 \times 10^{-5}$ \cite{Will} is well satisfied 
around $r=r_V$.
 
This argument shows that, as long as $r_V$ is larger than 
$r_{\oplus} \simeq 10^{15}$\,cm, the Vainshtein mechanism 
suppresses the fifth force inside the Solar System.
On using Eq.~(\ref{rVes2}) for the Sun with 
$H_0^{-1} \simeq 10^{28}$\,cm, 
this condition translates to $\xi_M \lesssim 10^{16}\,|q|$.
On the other hand, the today's value of $x_4$ defined in Eq.~(\ref{varidef}) 
is related to $x_1$, as $x_4(0)=15x_1^4(0)/[\xi_M^2(1-s)F]$. 
As we see in Figs.~\ref{xifig} and \ref{xifig2}, the value of 
$x_1^2$ at $z=0$ is at most of the order $10^{-2}$. 
Then, the cosmological bound (\ref{x4con1}) coming from 
a viable background expansion history translates to 
$\xi_M \gtrsim 10$ for $(1-s)F=O(1)$.
Hence the mass scale $M$ consistent with both local gravity 
and cosmological constraints is given by  
\be
10 \lesssim \xi_M \lesssim 10^{16}\,|q|\,,
\ee
whose allowed range is broad.

Finally, around the center of a spherically symmetric body ($r<r_g$), 
there is a regular solution 
\be
\phi'(r)={\cal C}r\,.
\label{phias}
\ee
The constant ${\cal C}$ is known by substituting Eq.~(\ref{phias}) 
into Eq.~(\ref{phieq}) with 
$\mu_1 \simeq -qr/(2\beta M_{\rm pl})$ and 
$\mu_2 \simeq -[a_1r/2+24(a_4-b_4){\phi'}^3/r^2]/\beta$, i.e.,
\be
\frac{{\cal C}^3}{3M^6}+\omega {\cal C}
-\frac13 a_1=\frac{q\rho_m}{3M_{\rm pl}}\,.
\label{Ceq}
\ee
Under the operation of the Vainshtein mechanism, the first term 
in Eq.~(\ref{Ceq}) is the dominant contribution to the lhs, so that 
${\cal C} \simeq (q\rho_m M^6/M_{\rm pl})^{1/3}$.
Using the solution (\ref{phias}), the parameter 
$\alpha_{\rm H}=-a_4(1-3s){\phi'}^4/A_4$ vanishes 
in the limit that $r \to 0$. 
Hence there is no conical singularity at $r=0$ in our model.
 
%===========================================
%section V
\section{Conclusions}
\label{consec} 
%===========================================

In the framework of GLPV theories, we have proposed a dark energy 
model in which the Vainshtein mechanism is at work without 
a conical singularity at the center of a compact body. 
In the presence of a diatonic coupling $F(\phi)=e^{-2q\phi/M_{\rm pl}}$, 
the field self-interactions $a_4X^2$ and $b_4X^2$ appearing 
in the functions $A_4$ and $B_4$, which exhibit the deviation from 
Horndeski theories for $a_4 \neq 3b_4$, can lead to the recovery 
of GR inside the Solar System. 
We have considered a massless scalar field with the 
kinetic term $-(1-6q^2)F(\phi)X/2$,
such that Brans-Dicke theory is encompassed as a specific case. 
In the limits that $q \to 0$ and $a_4 \to 3b_4$, our theories also 
recover a sub-class of covariant Galileons.

We studied the background cosmology for the linear 
potential $V(\phi)=a_1\phi$ (which is responsible 
for the late-time cosmic acceleration).
The coupling $q$ can induce 
a $\phi$-matter-dominated-epoch, but the period of 
$\phi$MDE depends on the amplitude of nonlinear 
field self-interactions. If the variable
$x_4=10a_4\dot{\phi}^4/(M_{\rm pl}^2F)$ is not 
initially very large, the growth of $x_4$ ends before 
the onset of $\phi$MDE (see Fig.~\ref{xifig}).
For larger $x_4$, the field self-interaction term dominates 
over the standard field kinetic energy during the deep matter
era (see Fig.~\ref{xifig2}). In the latter case, the realization 
of a proper matter era requires that 
the today's value of $x_4$ is in the range 
$x_4(z=0) \lesssim 10^{-6}$ for $|q|< O(0.1)$.

In GLPV theories the scalar propagation speed $c_{\rm s}$ 
has a non-trivial mixing with the matter sound speed. 
Under the condition that $x_4$ dominates over other kinetic 
contributions in the early cosmological epoch, we derived the analytic 
formulas of $c_{\rm s}^2$ during the radiation and matter eras, 
see Eqs.~(\ref{csrad}) and (\ref{csmat}) respectively. 
They are in good agreement with the numerical 
results shown in Figs.~\ref{csfig} and \ref{csfig2}, 
where $c_{\rm s}^2$ finally approaches the value 1 
after the suppression of nonlinear field self-interactions. 
We find that, as long as the ratio $s=b_4/a_4$ is in the range 
$0 \le s \le 1$, the Laplacian instability associated with negative 
values of $c_{\rm s}^2$ can be avoided. 
The deviation of the tensor propagation speed squared 
$c_{\rm t}^2$ from 1 is as small as $x_4$, so there is 
no instability for the tensor mode.

Since $c_{\rm s}^2$ is at most of the order of 1, our theory is 
not plagued by the existence of heavy oscillating modes of 
perturbations. In fact, we have numerically confirmed that 
the gauge-invariant gravitational potentials $\Psi$ and $\Phi$
exhibit regular behavior without heavy oscillations 
(see Fig.~\ref{grafig}). The difference between $-\Psi$ and 
$\Phi$ is given by Eq.~(\ref{PsiPhi}), so 
the gravitational slip parameter $\gamma=-\Phi/\Psi$ is 
close to 1 for $qx_1$ and $x_4$ much smaller than 1. 
We also studied the growth of 
matter perturbations $\delta_m$ and found that the 
effective gravitational coupling is well described by 
$G_{\rm eff}\simeq (G/F)(1+2q^2)$ in the late 
cosmological epoch. The current measurements of 
redshift-space distortions alone are not accurate enough to 
distinguish between the models with different 
values of $q$ (see Fig.~\ref{fsigmafig}).

On the spherically symmetric background with a matter source, 
we also derived the field profile and the Vainshtein radius $r_V$.
The resulting solutions to the two gravitational potentials 
are given by Eqs.~(\ref{Phiso}) and (\ref{Psiso}) in 
the regime $r<r_V$.
We showed that, as long as $r_V$ is larger than the solar-system 
scale $r_{\oplus} \simeq 10^{15}$\,cm, the gravitational 
slip parameter $\gamma$ is within the  
upper bound of local gravity experiments.
For the consistency with both cosmological and local gravity 
constraints, the mass scale $M=[24a_4(1-s)]^{-1/6}$ and 
the ratio $s$ are bounded to be in the ranges 
$10 \lesssim M^3/(H_0^2M_{\rm pl}) \lesssim 10^{16}|q|$
and $0 \le s<1$, respectively.

It is of interest to place more precise bounds on the 
parameters $M$, $s$, and the coupling $q$ by employing the 
CMB data combined with other observational data (along the line 
of Refs.~\cite{mcamb}). This is left for a future work.

%%%%%%%%%%%%%%%%%%%%%%%%%%%%%%%%%%%%
\section*{ACKNOWLEDGEMENTS}
RK is supported by the Grant-in-Aid for Research Activity Start-up 
of the JSPS No.\,15H06635. 
ST is supported by the Grant-in-Aid for Scientific Research Fund of
the JSPS No.\,24540286, MEXT KAKENHI Grant-in-Aid for Scientific Research 
on Innovative Areas ``Cosmic Acceleration'' (No.\,15H05890),  
and the cooperation program between Tokyo
University of Science and CSIC.
%%%%%%%%%%%%%%%%%%%%%%%%%%%%%%%%%%%%

%===============================================================%
%************************ APPENDIX ****************************%
%===============================================================%

\end{document}